\documentclass[preprint,showpacs,showkeys,aps,prd,amsfonts,eqsecnum,nofootinbib]{revtex4}
\usepackage{graphicx,epsfig}
\usepackage[dvips]{color}
\usepackage{amsmath, amssymb, graphics}
\def\gsim{\lower0.5ex\hbox{$\:\buildrel >\over\sim\:$}}
\def\lsim{\lower0.5ex\hbox{$\:\buildrel <\over\sim\:$}}
\begin{document}
\preprint{CUMQ/HEP 143}
%
%
\title{\Large  CP Asymmetry in Charged Higgs Decays in MSSM}
\author{Mariana Frank}\email[]{mfrank@alcor.concordia.ca}
\author{Ismail Turan}\email[]{ituran@physics.concordia.ca}
\affiliation{Department of Physics, Concordia University, 7141
Sherbrooke Street West, Montreal, Quebec, CANADA H4B 1R6}
\date{\today}

\begin{abstract}
We discuss and compare the charge-parity (CP) asymmetry in the charged Higgs boson decays 
$H^-\to \bar{u}_i d_j$ for the second and third generation quarks in the minimal supersymmetric standard model. As part of the analysis, we derive some general analytical formulas for the imaginary parts of two-point and three-point scalar one-loop integrals and use them for calculating vectorial and tensorial type integrals needed for the problem under consideration. We find that, even though each decay mode has a potential to yield a CP asymmetry larger than $10\%$, further analysis based on the number of required charged Higgs events at colliders favors the $\bar{t}b$, $\bar{c}b$, and $\bar{c}s$ channels, whose asymmetry could reach $10-15\%$ in certain parts of the parameter space. 

\end{abstract}
\maketitle
\section{Introduction}\label{sec:intro}
Discrete charge-parity (CP) symmetry 
violation is an interesting and puzzling phenomenon within the standard model (SM). While the SM has been successful in explaining all of the available experimental data, it fails to provide an explanation for the breaking of CP. Insight into  
understanding its nature and structure would shed light on diverse issues ranging from the origin 
of the mass to the evolution of our universe.

The SM presents an economical scenario for CP violation, described through  only one weak phase in Cabibbo-Kobayashi-Maskawa (CKM) matrix. 
Even though the observed direct CP violation in K \cite{Yao:2006px} and B \cite{Aubert:2001fg} decays can be 
accommodated via the CKM matrix of the SM, this is not true for other phenomena, such as the matter-antimatter
asymmetry  present in the universe. In addition, the Standard Model fails to explain the origin of CP violation but merely parameterizes it. Thus the SM framework could be viewed as a low energy effective version of a more complete 
theory, and 
CP violation offers a motivation to go beyond the SM. In the models beyond the SM, the existence of new sources of CP violation other than the CKM phase,
 resolves some of the problems of the SM, but not all. One of the leading candidates of physics beyond the SM 
is supersymmetry,  in particular the minimal supersymmetric standard model (MSSM). Supersymmetry provides a compelling argument for the stabilization of the Higgs sector against quadratic divergencies and allows unification of gauge couplings at high energies, both unexplained by the SM. 
In the MSSM, enough baryon asymmetry can be generated at the electroweak scale at any temperature 
by means of the existence of an additional Higgs doublet.  However, the MSSM has difficulty reconciling the smallness of the electric dipole moments with expectations of scalar fermion masses and the size of new CP violating phases introduced by soft supersymmetry breaking  \cite{Buchmuller:1982ye}.

In MSSM, there are many parameters which can in principle be complex, even after making all 
allowed rotations to get rid of unphysical phases. This raises the interesting possibility that CP violation, or CP asymmetries, might arise in other sectors of the model than the quark sector, and that would provide a spectacular signal of physics beyond the standard model. In particular,  it is possible that there might be a close relationship between the Higgs sector and CP violation. The structure and properties of the Higgs sector are under intense scrutiny at present. Though indications at LEP for a light Higgs boson of mass around 115 GeV are encouraging, they await confirmation.

Searching for Higgs bosons is one of the major objectives
of present and future high energy experiments. In particular, a charged Higgs boson, 
predicted by most models to have mass of the order of the weak scale, would be a definite sign for physics beyond the SM. 
The phenomenology of charged Higgs bosons is less model dependent than that of neutral Higgs bosons, and it is governed by the values of $\tan \beta$ and $m_{H^{\pm}}$. Because charged Higgs couplings are proportional to fermion masses, the decays to third generation quarks and leptons are dominant. At hadron
colliders, such as the Fermilab Tevatron and the CERN Large Hadron
Collider (LHC), a light charged Higgs boson can be produced from
the decay of top quark via $ t \to H^+ b$, if
$m_{H^\pm}<m_t-m_b$. If the charged Higgs
boson is heavier than the top quark, there are three channels for producing charged Higgs pairs: $pp \to H^+H^-,~pp \to W^{\pm}H^{\mp}$ and 
$gb \to tH^-$;  as well as the single charged Higgs production $\bar cs,\bar cb \to H^-$; see \cite{Hao:2007qv} and references therein. In many cases complementary  information from more than one channel will be accessible at the LHC. The LHC has a high potential for detecting heavy Higgs states which might be beyond the kinematic reach of the planned International Linear Collider (ILC). 
 At the ILC, the main production mechanism for charged Higgs bosons would be $e^+e^- \to H^+H^-$, followed by one of the allowed decays $H^+ \to t\bar{b}, \tau \nu_{\tau}$ or $c \bar{s}$. The pair production cross section for charged Higgs at ILC is about 2.5 larger than for the neutral ones \cite{Coniavitis:2007me}.
Provided that a Higgs boson couples to the Z boson, the ILC will observe it independently of its decay characteristics.  The discovery potential is practically independent of $\tan \beta$ and extends up to 1.2 TeV for an integrated luminosity of 3000 fb$^{-1}$\cite{Coniavitis:2007me}.  For specific examples of how the integrated information obtained  by ILC and LHC can be used for Higgs detection and determination of parameters in the Higgs sector, see \cite{Weiglein:2004hn}.

In its most general form, the MSSM predicts a plethora of new CP phases. These new sources of flavor and CP violation give rise to the enhancement of CP violation effects alluded to before, which could provide distinguishing signs for MSSM at present and future colliders. In this study, we concentrate on analysis of the CP asymmetry in charged Higgs decays $H^-\to\bar{u}_i d_j$ in
the framework of the MSSM. Here $\bar{u}_i d_j = \bar{c}b, \bar{c}s, \bar{t}b, \bar{t}s$. The corresponding neutral Higgs CP asymmetry in $h\to d_i\bar{d}_j$ has been discussed in Ref.~\cite{Demir:2003bv}. The CP asymmetry in the main decay mode $H^-\to\bar{t}b$ and other two-body non-quark charged Higgs decays have been considered in \cite{Christova:2002ke,Christova:2002sw}. Recently, these authors have discussed various CP asymmetries in $H^-\to\bar{t}b$ by including the decay products of the top quark and subsequently the W boson and showed that the decay rate CP asymmetry can go up to $25\%$ \cite{Christova:2006fb}.  In this work, we revisit the asymmetries in $H^-\to\bar{t}b$, but also discuss the other three quark decay modes and compare the  size of the CP asymmetry in all the channels. We 
show that even though in some part of the parameter space, the $\bar{t}s$ channel has sizable CP asymmetry with respect to the $\bar{t}b$, $\bar{c}b$, or $\bar{c}s$ channel, 
this result has to be taken with caution since the former has very small branching ratios ($Br$'s) 
which makes harder to observe it. We introduce and discuss another 
relevant quantity\footnote{$(A_{CP}^2 \times Br)^{-1}$ is closely related to the total number of events $N$ required to establish a measurable CP violation for a particular mode. The exact formula is $N=s^2(A_{CP}^2 \times Br\,\epsilon)^{-1}$. Here $s$ is the standard deviation and $\epsilon$ is the detection efficiency.} \cite{Eilam:1991yv}, $(A_{CP}^2 \times Br)^{-1}$, for each decay mode,  with $A_{CP}$  the CP asymmetry, and $Br$ the branching ratio for a given decay mode. This function was shown to be a measure of the  number of 
required charged Higgs bosons to be produced at colliders for observing an asymmetry for a given channel. Based on this analysis, we conclude that  $H^-\to \bar{c}b$, $H^-\to \bar{c}s$,  and $H^-\to \bar{t}b$ are all  
optimal channels which could reveal a 
measurable CP asymmetry at the order of $10 - 15\%$. We also discuss the CP asymmetry induced by 
the phases of the flavor violating parameters $\delta_{U,D}^{23}$ alone and note that their contribution is small and thus  unlikely to account for a 
measurable CP asymmetry in any of the decay modes considered in this study.

The outline of the paper is as follows. In the next section, we briefly review the MSSM, concentrating on the sources of CP violation. The decay 
processes are presented in Section \ref{decays} and the numerical analysis of the decays under consideration in the Section \ref{numerical}. We conclude 
in Section \ref{conc}. Details of the method used for calculations are presented in the Appendix.

\section{The Unconstrained Minimal Supersymmetric Standard Model}
\label{model}

The superpotential $\mathcal{W}$ of the MSSM Lagrangian  and the relevant 
part of the soft breaking Lagrangian 
$\mathcal{L}^{\text{squark}}_{\text{soft}}$ are respectively 
\begin{eqnarray}
     \label{eq:W} 
\!\!\!\!\!\!\!\!{\mathcal{W}} &=& \mu H^1 H^2 + Y_l^{ij} H^1
{L}^i {e}_R^j + Y_d^{ij} H^1 {Q}^i {d}_R^j
+ Y_u^{ij} H^2 {Q}^i {u}_R^j\\
\!\!\!\!\!\!\!\!\!\!\!\!\!\!\mathcal{L}^{\text{squark}}_{\text{soft}}\!\!\! &=&\!\!\!
-\tilde Q^{i\dagger} (M_{\tilde Q}^2)_{ij} \tilde Q^j
-\tilde u^{i\dagger} (M_{\tilde U}^2)_{ij} \tilde u^j
-\tilde d^{i\dagger} (M_{\tilde D}^2)_{ij} \tilde d^j 
+ Y_u^i A_u^{ij} \tilde Q_i H^2 \tilde u_j
+ Y_d^i A_d^{ij} \tilde Q_i H^1 \tilde d_j,
\label{eq:superpot}
\end{eqnarray}
where $H^1$ and $H^2$ are the Higgs doublets with vacuum expectation values 
$v_1$ and $v_2$ respectively, $ Q$ is the $SU(2)$ scalar 
doublet, $ u$, $ d$ are 
the up- and down-quark $SU(2)$ singlets, respectively,  $\tilde Q, \tilde u, \tilde d$
represent scalar quarks,  $Y_{u,d}$ are the
Yukawa couplings and $i,j$ are generation indices. Here $A^{ij}$ represent the 
trilinear scalar couplings. In Eq.~(\ref{eq:superpot}) we are assuming a chiral limit of MSSM. 

We work in the unconstrained version of the MSSM and use the mass eigenstate method \cite{Besmer:2001cj}, where squark mass matrices are given in the super-CKM basis, and are diagonalized by rotating the superfields.  In this basis, the up-squark and down-squark mass matrices are correlated by this rotation and thus not independent. Potential new sources of flavor violation arise from couplings of quarks and squarks to gauginos. This method has the advantage that, when the off-diagonal elements in the squark mass matrices become large, the method is still valid, unlike perturbation-based expansions. The up(down)-squark mass matrices between second and third generations are taken as
\begin{equation}
\label{eq:squarkmass}
\!\!\!\!\!\!\!\!{\cal M}^2_{\tilde {u}\{\tilde d\} }=
\left( \begin{array}{cccc}
M_{{\tilde L} c\{s\}}^2 & (M^2_{\tilde U\{\tilde D\}})_{LL} & m_{c\{s\}} {\cal A}_{c\{s\}}
&(M^2_{\tilde U\{\tilde D\}})_{LR} \\
(M^2_{\tilde U\{\tilde D\}})_{LL} & M_{{\tilde L} t\{b\}}^2 &
(M^2_{\tilde U\{\tilde D\}})_{RL} & m_{t\{b\}} {\cal A}_{t\{b\}} \\[.3ex]
m_{c\{s\}} {\cal A}_{c\{s\}} & (M^2_{\tilde U\{\tilde D\}})_{RL} &M_{{\tilde R} c\{s\}}^2 &
(M^2_{\tilde U\{\tilde D\}})_{RR} \\
(M^2_{\tilde U\{\tilde D\}})_{LR} & m_{t \{b\}} {\cal A}_{t \{b\}} &
(M^2_{\tilde U \{\tilde D\}})_{RR} &M_{{\tilde R} t \{b\}}^2
\end{array} \right)
\end{equation}
with
\begin{eqnarray}
\label{eq:squarkparam}
M_{{\tilde L}q}^2 &=&
      M_{\tilde Q,q}^2 + m_q^2 + \cos2\beta (T_q - Q_q s_W^2) M_Z^2\,,\;
\nonumber \\
M_{{\tilde R}\{c,t\}}^2 &=&
      M_{\tilde U,\{c,t\}}^2 + m_{c,t}^2 + \cos2\beta Q_t s_W^2
M_Z^2\,, \nonumber \\
M_{{\tilde R}\{s,b\}}^2 &=&
      M_{\tilde D,\{s,b\}}^2 + m_{s,b}^2 + \cos2\beta Q_b s_W^2 M_Z^2\,,\\
{\cal A}_{c,t} &=& A_{c,t} - \mu\cot\beta\,,\;\;\;
{\cal A}_{s,b} = A_{s,b} - \mu\tan\beta\,. \nonumber
\end{eqnarray}
where we assume a general flavor violation among families. In addition to the flavor dependence in Yukawa matrices, there are additional sources of flavor violation due to the soft mass terms and $A$-terms. The richness of the flavor structure depends on the assumed textures of soft mass and $A$ terms at the GUT scale.  Assuming both the soft mass and $A$-terms to be flavor blind at GUT scale can still induce flavor violation at electroweak scale due to their evolution from GUT scale down to the electroweak scale using the renormalization group equations. The flavor violating effects become much bigger if the soft mass and/or $A$-term(s) are flavor dependent at GUT scale (see, for example, \cite{Chankowski:2005jh} for details).

For reasons we discuss further on, we assume $\mu$ real and a common phase for $A_{c,t}$ and $A_{s,b}$. Note that 
$A_s$ has negligible effect since it is multiplied by the strange quark mass. $\tan\beta$ is the ratio of the vacuum expectation values of the two neutral Higgs bosons.

 From the mass matrix, Eq. (\ref{eq:squarkmass}), we also define the flavor mixing parameters as scaled off-diagonal flavor violating entries
\begin{eqnarray}
(\delta_{U(D)}^{ij})_{AB} =
\frac{(M^2_{\tilde{U}(\tilde{D})})_{AB}^{23}}{M^2_{\rm{SUSY}}},\;\; 
(A,B=L,R),
 \label{deltadefb}
\end{eqnarray}
where $M_{\rm SUSY}$ is the common scale for the parameters 
$M^2_{\tilde{Q},q}$ and $M^2_{\tilde{U}(\tilde{D}),q}$. As mentioned before 
 we allow $\delta_{U(D)}^{23}$'s  to be complex (to have CP violating phases).

We do not repeat listing the chargino and neutralino sectors of the MSSM here, 
as we don't assume a new non-zero  CP phase in either sector;  for the details 
see Ref.~\cite{Frank:2006ku}. As the gluino contribution is dominant
for the charged Higgs decays, we give the relevant up-type quark-squark-gluino interaction 
$\tilde{g}$:
\begin{equation}
\mathcal{L}_{u \tilde{u} \tilde g}= \sum_{i=1}^{3}\sqrt{2}\, g_s \,
T^r_{st} \left[ \bar
u^{s}_i \,(\Gamma_U)^{ia}\,P_L\, \tilde g^r \,\tilde u^{t}_a - \bar
u^{s}_i \,(\Gamma_U)^{(i+3)a}\,P_R \,\tilde g^r \,\tilde u^{t}_a +
\text{H.c.} \right]\,,
\end{equation}
where $T^{r}$ are the $SU(3)_{c}$ generators,  $P_{L,R}\equiv (1\mp
\gamma_5)/2$, $i=1,2,3$ is the generation index, $a=1, \ldots, 6$ is
the scalar quark index, and $s,t$ are color indices. There is a similar interaction for the down case. 

Considering the number of complex parameters in MSSM, further assumptions are needed for simplicity and predictability. 
In the gaugino sector, two of three gaugino masses could be complex, unless a degenerate spectrum at 
grand unified theory (GUT) scale is assumed. We assume all three gaugino
mass terms real at any scale. The higgsino mass parameter $\mu$ is in principle complex. However, 
the phase of $\mu$ is strongly constrained by the electron dipole moment (EDM) measurements of neutron 
\cite{Fischler:1992ha} and cannot exceed values of the order of 0.01-0.001. For the effects of CP violating SUSY phases on other EDMs and systems, see \cite{Demir:2003js}. We simply neglect this phase and 
 consider $\mu$ real. In the squark sector, there are trilinear soft couplings of quarks, $A_{u,d}$,
which are complex. In addition to these, the misalignment between quarks and squarks arising 
through the diagonalization of their respective mass matrices leads to new sources of flavor violation, with parameters 
denoted as $\delta_{U,D}$. These are also generally complex, and could have large imaginary 
parts. We will assume non-zero flavor violation only between the second and third generations, 
 because the ones involving the first generation are required to be small, based on experimental constraints in $K$ and  
$D$ physics \cite{Chankowski:2005jh,Gabbiani:1996hi}. The kaon mass splitting parameter $\Delta M_K$ due to $K-\bar{K}$ mixing, and the parameters $\epsilon$ and $\epsilon^\prime$ put severe constraints on the real and imaginary parts of the flavor violating parameters $\delta^{12}_{D}$, respectively. Somewhat weaker bounds can also be obtained from K physics for the up sector between the first and second generations as well but stringent bounds on $\delta^{12}_{U}$ can be obtained by using the experimental bound on the $D-\bar{D}$ mixing parameter $\Delta M_D$. In a similar fashion, the flavor violating parameters $\delta^{13}_{D}$ between the first and third generations can be restricted with the use of low energy B physics measurements ($\Delta M_{B_d},B\to X_s \gamma, S_{B\to\psi K_s}$). The common feature of these constraints is that the upper bounds on the flavor parameters involving the first generation have to be  less than 0.1 \cite{Chankowski:2005jh,Gabbiani:1996hi}. There are however no similar limits on the mixings between the second and third generations so that we will keep them arbitrary. 

We assume that there are non-zero phases from the trilinear 
couplings $A_{u,d}$ and intergenerational flavor mixing parameters $\delta_{U,D}^{23}$. For further 
simplification, we assume a common phase ${\rm Arg}[A_u] = {\rm Arg}[A_d] \equiv {\rm Arg}[A]$. 

Thus, the supersymmetric sources of CP violation of interest in charged Higgs decays come from the soft broken
terms $M^2_{\tilde Q}$, $M^2_{\tilde U}$ and the trilinear scalar coupling $A_{u}$, and are
introduced through the matrix $\Gamma_U$. In the following section, we analyze their effects on the calculation of the CP asymmetry.
\section{Charged Higgs decays $H^-\to \bar{u}_i d_j$}
\label{decays}
In this section we discuss the CP asymmetry in the charged Higgs boson decays $H^-\to \bar{u}_i d_j$, which is defined as\footnote{This type of CP asymmetry is sometimes called {\it partial rate asymmetry}. For other types see \cite{Atwood:2000tu,Christova:2006fb}.}
\begin{equation}
\displaystyle A_{CP} =\frac{\Gamma(H^-\to \bar{u}_i d_j)-\Gamma(H^+ \to u_i \bar{d}_j)}{\Gamma(H^-\to \bar{u}_i d_j)+\Gamma(H^+ \to u_i \bar{d}_j)},
\label{ACP} 
\end{equation}
where $\Gamma$ is the partial decay width of the decay mode considered. These decays 
are tree level processes and one could calculate the branching ratios, decays widths,  
etc. with tree-level approximation.
\begin{figure}[htb]
\vspace*{-2.1in}
        \centerline{\epsfxsize 6.5in {\epsfbox{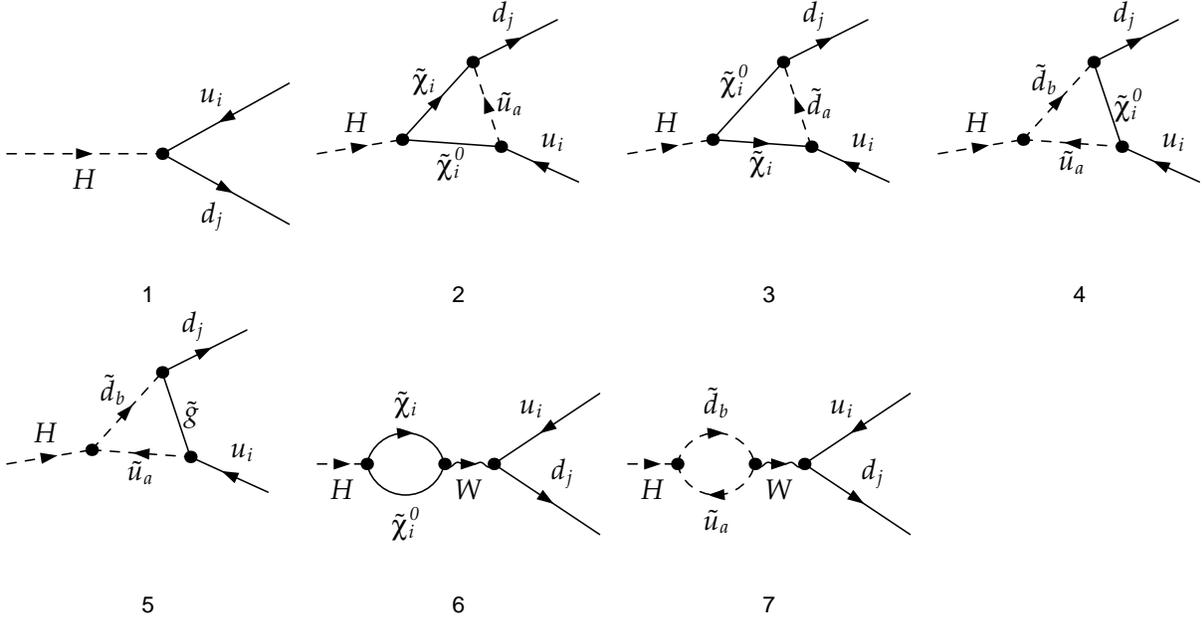}}}
\vspace*{-2.35in}
\caption{The tree and relevant one-loop diagrams contributing CP asymmetry for the decays $H^-\to \bar{u}_i d_j$.}
\label{fig:relevant}
\end{figure}

However, it is known that \cite{Atwood:2000tu} the CP-odd observable $A_{CP}$ requires a nontrivial phase from Feynman diagrams (called absorptive or strong phase),  in addition to the weak phase mentioned in the previous section. This way, the imaginary part of the amplitude is non-zero, ${\mathcal Im}(\rm Amplitude)\ne 0$. One way of introducing such a phase is through one-loop Feynman diagrams, where some of the intermediate particles go on-shell. Then, the numerator of Eq.~(\ref{ACP}) will be proportional to the interference term between tree level and one-loop contributions. The tree level and relevant one-loop contributions to the decays $H^-\to \bar{u}_i d_j$ are shown in Fig~\ref{fig:relevant}. 

\begin{figure}[htb]
\hspace*{-0.3cm} 
\centerline{\epsfxsize 6.2in {\epsfbox{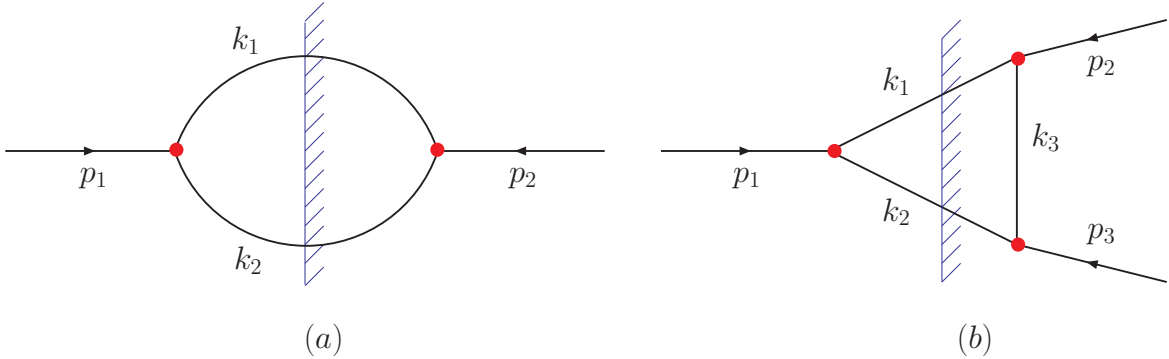}}}
\vspace*{-0.1in}
\caption{The unitarity cuts for self energy $(a)$ and vertex type $(b)$ generic diagrams.}
\label{fig:cuts}
\end{figure}

There are many more one-loop contributions to the decays, but based on various kinematical considerations only six of them are relevant to $A_{CP}$ (four vertex type and two self-energy type diagrams)\footnote{There exist additionally some SM vertex contributions with $W$ boson, CP-even Higgs bosons, and quarks in the loop. We don't get CP asymmetry contributions from such diagrams since we take $\mu$ real. For the case with complex $\mu$, see \cite{Christova:2002ke}.}. We consider cuts through  chargino-neutralino internal lines, or through up squark-down squark  lines. We will call them {\it internal-cut} states. They contribute to the CP asymmetry when $m_H\geqslant m_{\tilde{\chi}^0}+m_{\tilde{\chi}^+}$ and $m_H\geqslant m_{\tilde{u}}+m_{\tilde{d}}$ are satisfied, so that the chargino and the neutralino can go on-shell. This is the necessary condition to induce an absorptive part into the amplitude. The generic self-energy and vertex type diagrams with cuts are shown in Fig.~\ref{fig:cuts} where the {\it internal-cut} states represent two possibilities  in each case. Here we don't count the flipping cases ($k_1 \leftrightarrow k_2$), which are only relevant  to the vertex type diagrams, in which case the third intermediate state is different. For the vertex-type diagrams, Fig.~\ref{fig:cuts}(b), there are two other possible cuts, through $k_1-k_3$ or $k_2-k_3$. The cut with $k_1-k_3$ is not kinematically allowed since it requires $m_{d_j} \geqslant m_{\tilde{d}_a}+m_{\tilde{\chi}^0_n}$, $m_{d_j} \geqslant m_{\tilde{d}_a}+m_{\tilde{\chi}^+_c}$, or $m_{d_j} \geqslant m_{\tilde{d}_a}+m_{\tilde{g}}$. None of these conditions is possible. 

For the case with $k_2-k_3$ cut, based on the current experimental bounds on the mass of the lightest neutralino and the lightest up-type squarks, there could be some contributions to the decay mode $H^-\to \bar{t}d_j$ from the diagrams $(2)$ and $(4)$ of Fig.~\ref{fig:relevant} in a very narrow kinematical range ($m_{\tilde{u}}+m_{\tilde{\chi}^0} \lesssim m_t$). We simply neglect such contributions. There exists yet another way to produce the necessary absorptive cut, by taking the invariant mass squared for the final states, $(p_2+p_3)^2$, to be greater than $(m_{\tilde{\chi}^0}+m_{\tilde{\chi}^+})^2$ which results in some cuts in the phase space. This method was persued in \cite{Bi:1999is}, in the analysis of the three-body semileptonic top quark decays. 

One can show that the numerator of Eq.~(\ref{ACP}) is proportional to the imaginary part of the amplitude from loop diagrams\footnote{For example see \cite{Atwood:2000tu} for the details.} arising from tree level-loop interference terms. As usual,  we neglect possible loop-loop interference effects which are much smaller than tree-loop terms. In Appendix, instead of giving rather lengthy analytical results for this imaginary part, we  outline the method used to do the calculation using the cuts in Fig.~\ref{fig:cuts}. We present it in a generic way, valid for the four decay modes of the charged Higgs boson, $H^-\to \bar{u}_i d_j$.    

\section{Numerical Analysis}
\label{numerical}
In this section, we present comparatively our numerical results for the decays $H^-\to \bar{u_i}d_j\;(u_i=c,t,\; d_j=s,b)$\footnote{The CP asymmetry in one of these decays, $H^-\to\bar{t}b$ has been discussed in \cite{Christova:2002ke} under a different parameter set and assumptions. A good agreement is obtained once we switch our parameter values to their set.}. In the calculation of the CP asymmetry from the formula in Eq~(\ref{ACP}), we use the tree level values instead of one loop when evaluating the sum of the partial decay widths ($\Gamma$) in the denominator, which simplifies the numerical calculations. This approximation is not valid for the numerator where one loop contributions are needed to extract the CP phases.
\begin{figure}[htb]
\begin{center}$
	\begin{array}{cc}
\hspace*{-1.1cm}
	\includegraphics[width=3.6in]{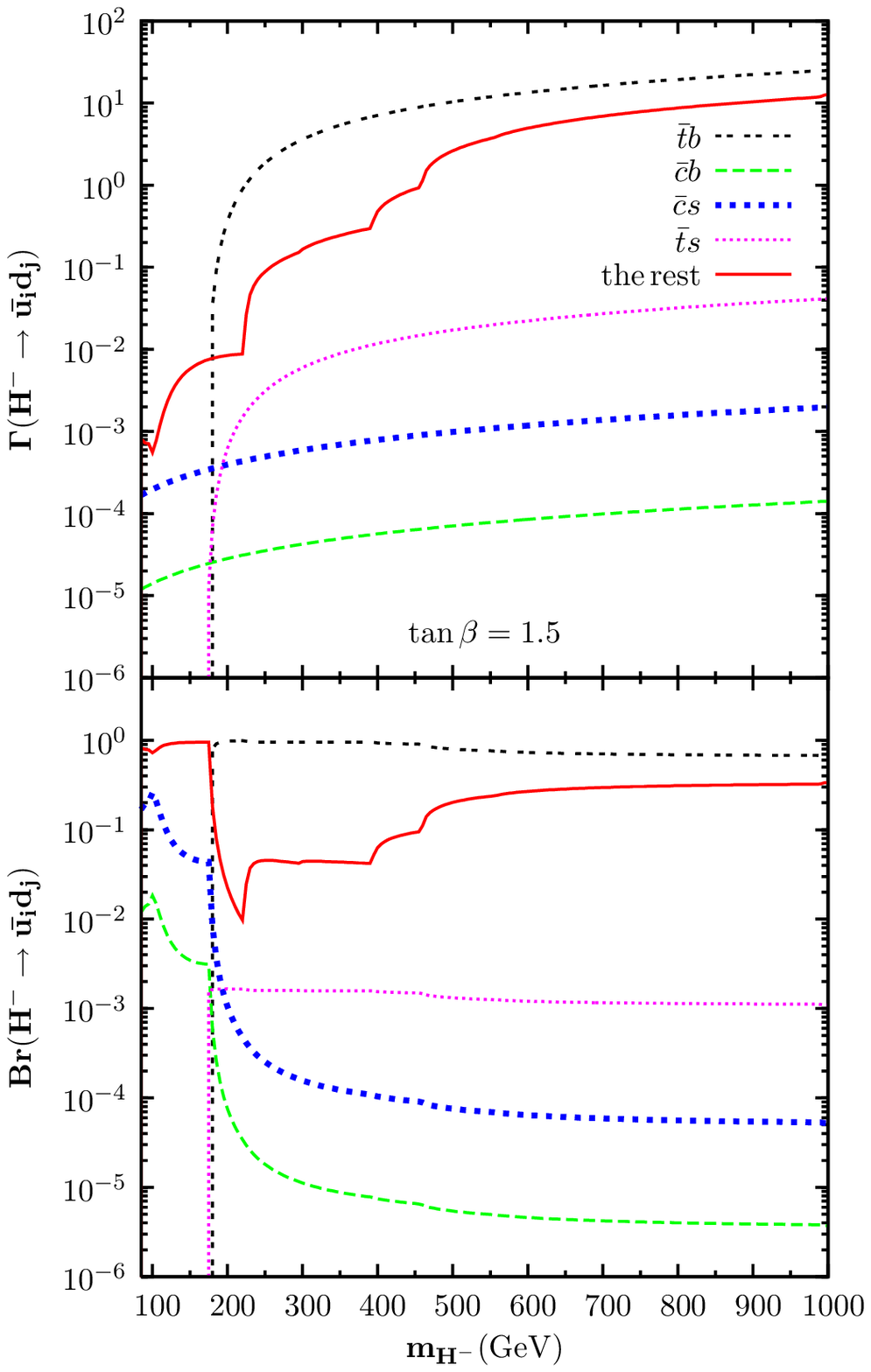} &\hspace*{-0.8cm}
	\includegraphics[width=3.6in]{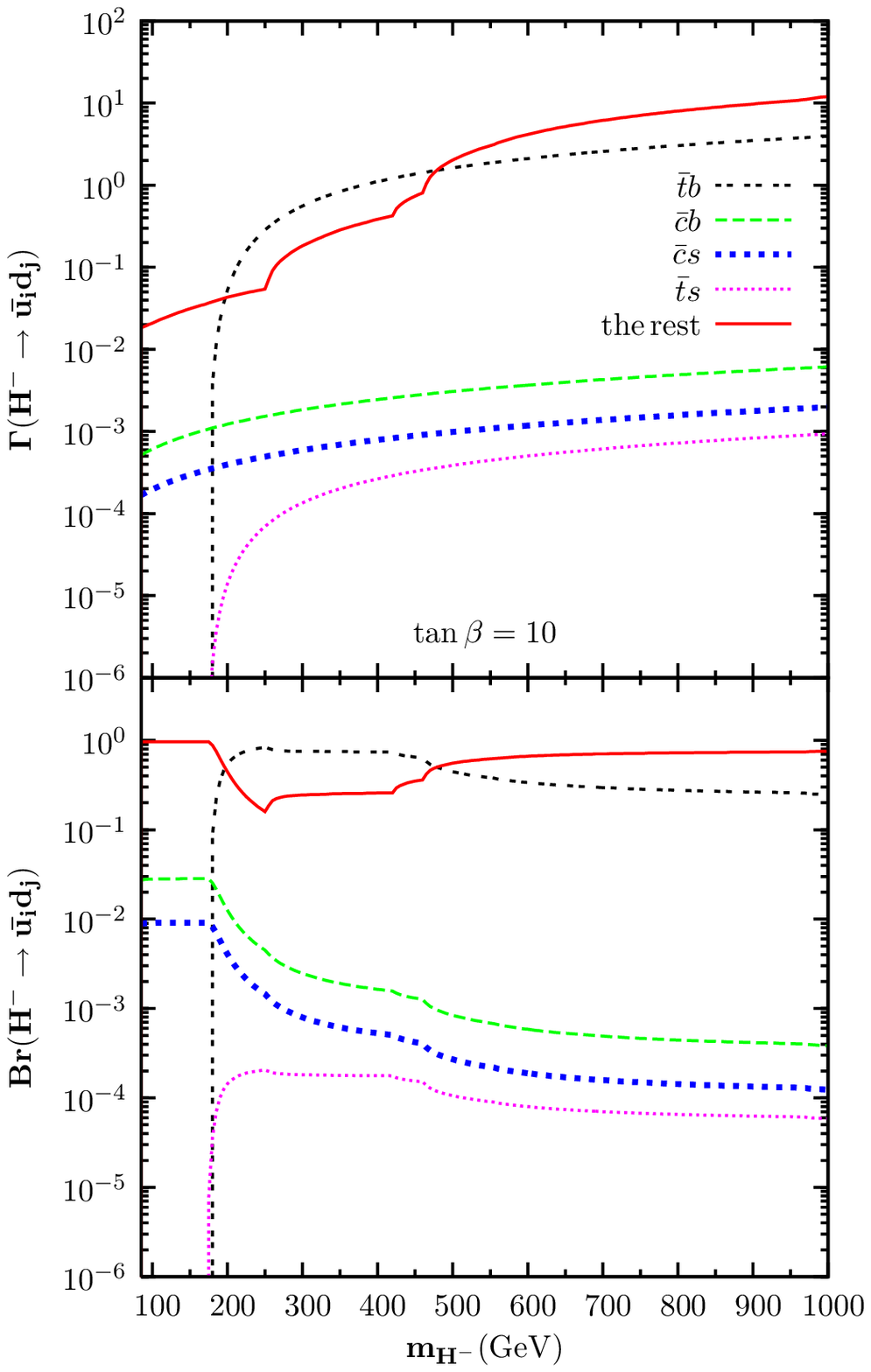}
	\end{array}$
\end{center}
\vskip -0.3in
      \caption{The partial decay widths and branching ratios of the charged Higgs boson decays $H^-\to \bar{u_i}d_j$. For the supersymmetric channels; $M_1=95$\, {\rm GeV}, $M_2=200$\, {\rm GeV}, $\mu=250$\, {\rm GeV}, $M_{\rm SUSY}=500$ GeV, and $A=400$\, GeV.}
\label{fig:Br}
\end{figure}

In Fig.~\ref{fig:Br}  we show the partial decay widths of the channels $H^-\to \bar{c}b,\,\bar{c}s,\,\bar{t}b,\,\bar{t}s$ at tree level, together with their branching fractions ($Br$) for small and intermediate $\tan\beta$ values. We represent the other tree level decay channels as ``{\it the rest}'' and we use the {\tt FeynHiggs} program \cite{FeynHiggs} to calculate the partial decay widths of these channels. Basically, depending on the charged Higgs mass, ``{\it the rest}'' includes the lepton channels $H^-\to e\nu_e,\,\mu\nu_\mu,\,\tau\nu_\tau$, the neutralino-chargino channels $H^+\to\chi^+_i\chi^0_j,\;i=1,2,\,j=1...4$, the Higgs-vector boson channels $H^\pm\to h^0W,\,H^0W,\,A^0W$, and the sfermion channels $H^\pm\to \tilde{f}_i\tilde{f}_j\;i,j=1...3$. Inclusion of these channels is important as it affects the number of charged Higgs bosons required to observe the asymmetry, as discussed later in this section.

In the region $m_{H^+} \geqslant m_t$, $H^-\to \bar{t}\, b$ is the dominant mode. However, in the region $m_{H^+} \leqslant 140$ GeV, the leptonic decay $H^\pm\to \tau\,\nu_\tau$ becomes the main decay channel, and in between these regions, including threshold effects, the below-threshold three body decay $H^\pm\to h\, W^*$ has a branching ratio comparable to, or even dominating over other channels, its exact value depending on the mixing in the Higgs sector. Inclusion of threshold effects also opens other three body channels, like $H^\pm\to A\, W^*$ and $H^-\to b\, \bar{t}^*$ that have sizable branching ratios in the intermediate mass range. We do not include the threshold effects, but see \cite{Djouadi:1995gv} for details. We note that for heavy charged Higgs, the neutralino-chargino channels are comparable with the $\bar{t}b$ channel. However, it is important to observe that among these decays, $H^-\to \bar{c}s$ and $H^-\to \bar{c}b$ have non-negligible branching ratios. This is important for our analysis, since observability of CP asymmetry in a specific channel requires not only a sizable asymmetry but also an experimentally viable branching ratio. We  must comment on the strange quark mass dependence: as seen from Fig.~\ref{fig:Br}, while the partial decay widths for $\bar{t}b$ and $\bar{t}s$ channels are suppressed as $\tan\beta$ gets larger, the opposite is true for the  $\bar{c}b$ channel. This is due to the fact that, evaluating the Feynman diagrams in Fig.~\ref{fig:relevant}, the scalar quark couplings to the charged Higgs are proportional to $( m_u \cot\beta +m_d \tan\beta)$, as seen from Eq.~(\ref{eq:squarkmass}). There is no supression in the $\bar{c}s$ curve since for intermediate $\tan\beta$ values the term proportional to the strange quark mass is comparable to the one with charm quark mass and becomes dominant for larger $\tan\beta$ values. So, one must keep the strange quark mass non-zero, at least for $\bar{c}s$ decay. Of course, its effect in the $\bar{t}s$ case is negligible.

For the numerical analysis,  we fix some of parameters of the model globally because the $CP$ asymmetry is not very sensitive to their variations. As mentioned before, we introduce a common phase and magnitude for the trilinear couplings $A_u$ and $A_d$ as ${\rm Arg[A]}$ and $A$. The Higgs parameter $\mu$ is taken real. We allow the gluino mass $m_{\tilde{g}}$ to be light. We also use the parameterization for squark mass matrices where a common scale $M_{\rm SUSY}$ is chosen for the soft breaking parameters $M^2_{\tilde{Q},q}$ and $M^2_{\tilde{U}(\tilde{D}),q}$. The flavor violating parameters ($\delta$'s ) are set to zero everywhere except for the case where we test the sensitivity of $A_{CP}$ to these parameters. There exist two other free parameters in the Higgs sector of the MSSM, taken in a popular  framework to be $\tan\beta$, and one of the Higgs boson masses, often taken as the $CP$-odd Higgs mass $m_A$. Of course we could equally well assume any of the others as the free Higgs mass parameter. As it is more convenient for the present analysis, we choose $m_{H^+}$ as the free Higgs mass here. Unless otherwise stated, we fix the following parameters globally\footnote{The gluino mass is consistent with TEVATRON limits \cite{Yao:2006px}} 
\begin{eqnarray}
M_{\rm SUSY} = 500\, {\rm GeV},\;\; \mu = 250\, {\rm GeV},\;\; \tan\beta= 10,\;\; M_2 = 200\, {\rm GeV},\;\; m_{\tilde{g}}= 250\, {\rm GeV}.
\end{eqnarray}
\begin{figure}[htb]
\begin{center}$
	\begin{array}{cc}
\hspace*{-0.5cm}
	\includegraphics[width=3.1in]{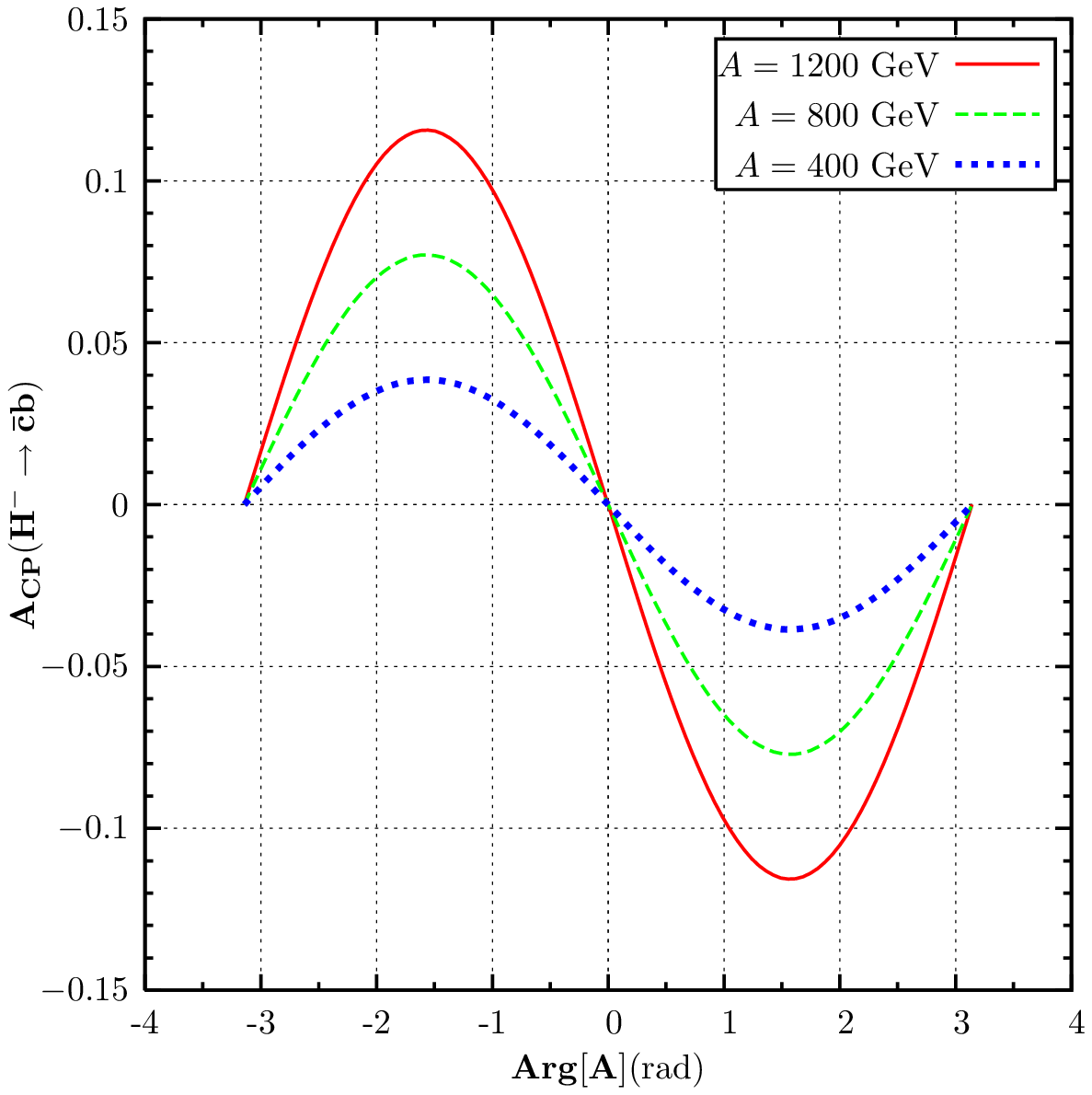} &
	\includegraphics[width=3.1in]{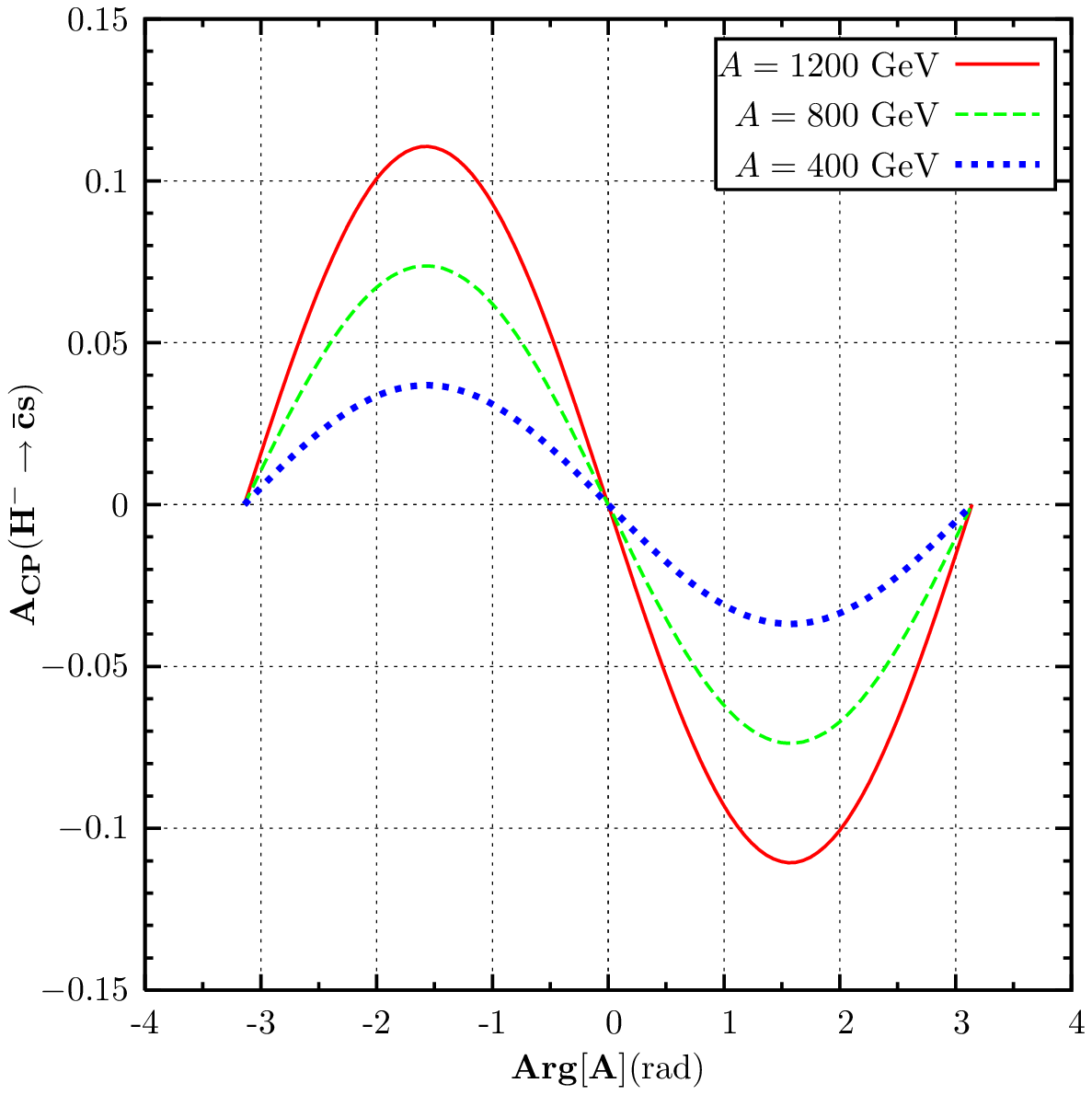} \\
\hspace*{-0.5cm}
	\includegraphics[width=3.1in]{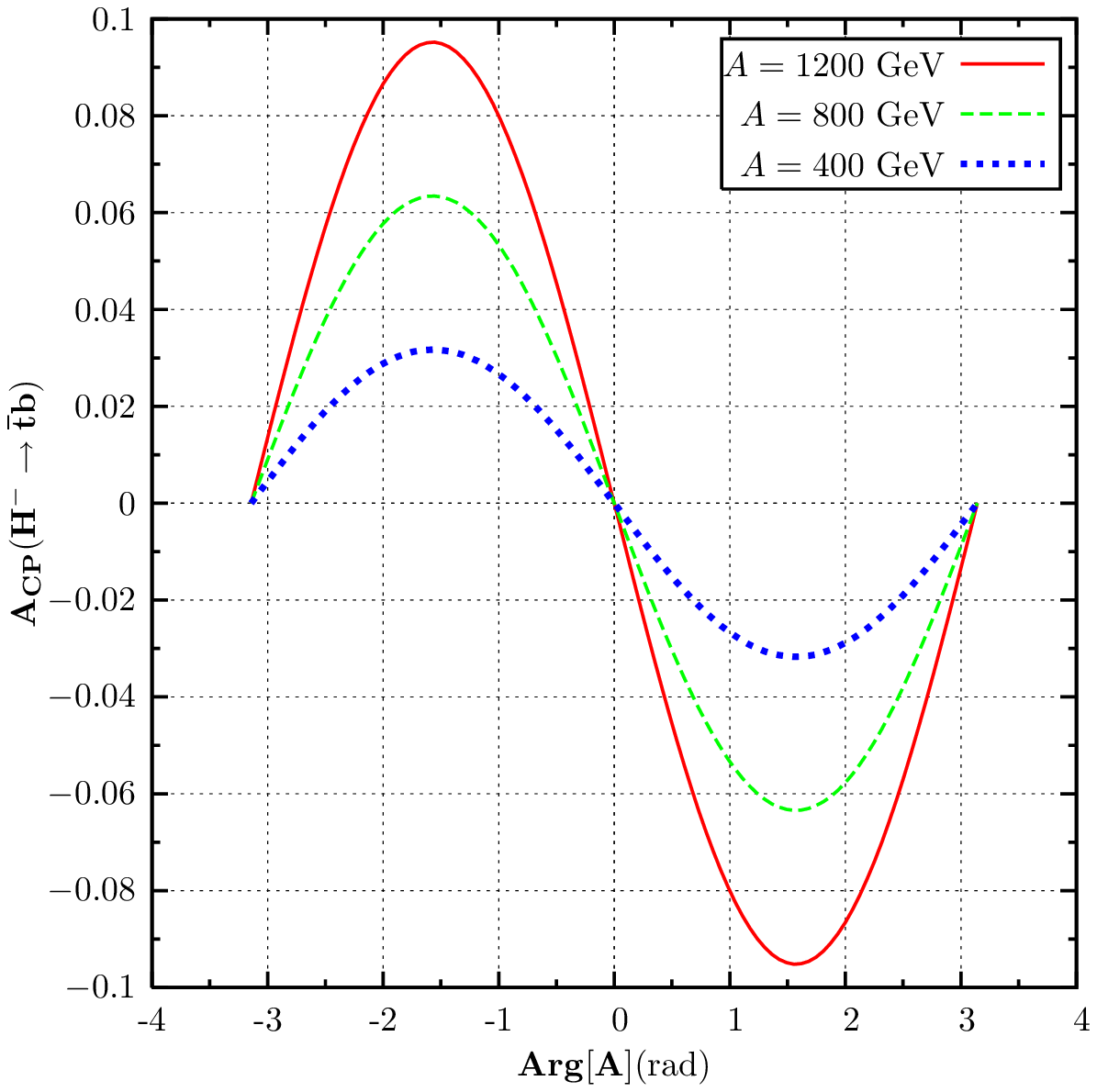} &
	\includegraphics[width=3.1in]{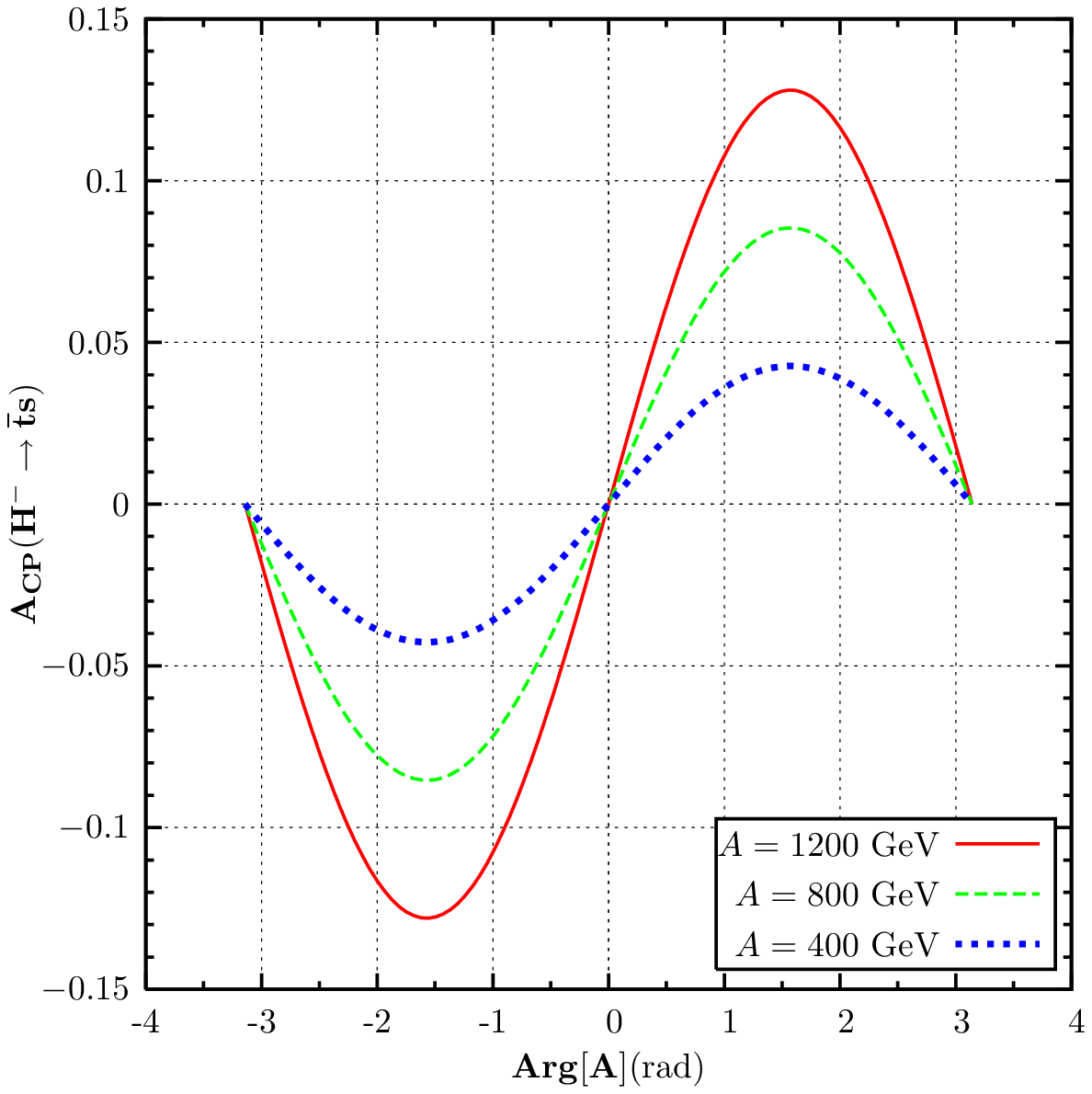}
	\end{array}$
\end{center}
\vskip -0.2in
      \caption{The $CP$ asymmetry for the charged Higgs decays $H^-\to \bar{u}_i d_j$ 
      as a function of the phase ${\rm Arg}[A]$ for the various values of $A$.}
\label{fig:phiA}
\end{figure}
In Fig.~\ref{fig:phiA}  we illustrate the dependence of the $CP$ asymmetry $A_{CP}$ on the phase ${\rm Arg}[A]$ for each decay for a variety of  $A$ values. As expected, the behaviour is periodic and the sensitivity to both ${\rm Arg}[A]$ and $A$ is quite significant. It seems that each decay mode can have an asymmetry as large as $10 -15\%$. It is also seen that the asymmetry produced in $H^-\to\bar{c}s$ and  $H^-\to\bar{t}s$ decay modes could be comparable, or even larger than $H^-\to\bar{c}b$ or $H^-\to\bar{t}b$. Of course this is not very unusual, since it is possible to get large asymmetries for decays with smaller branching ratios \cite{Atwood:2000tu}. Therefore, determining the optimum channel among these decays requires further analysis, but qualitatively one can predict that $H^-\to\bar{c}b$ and $H^-\to\bar{t}b$ have similar asymmetries. The $\bar{c}b$ mode has non negligible branching ratios due to the fact that CKM supression is compensated by the large $m_b$ mass appearing in the couplings. In the parameter space that we explore, the diagrams with chargino-neutralino/neutralino in the loop give negligible results compared to the gluino loop diagrams. So, the main contribution comes from gluino vertex diagram (diagram 5 of Fig.~\ref{fig:relevant}), while the self energy diagram (diagram 7 of Fig.~\ref{fig:relevant}) is also important.  We keep these contributions, but include and check the others wherever relevant.    
\begin{figure}[htb]
\begin{center}$
	\begin{array}{cc}
\hspace*{-0.5cm}
	\includegraphics[width=3.2in]{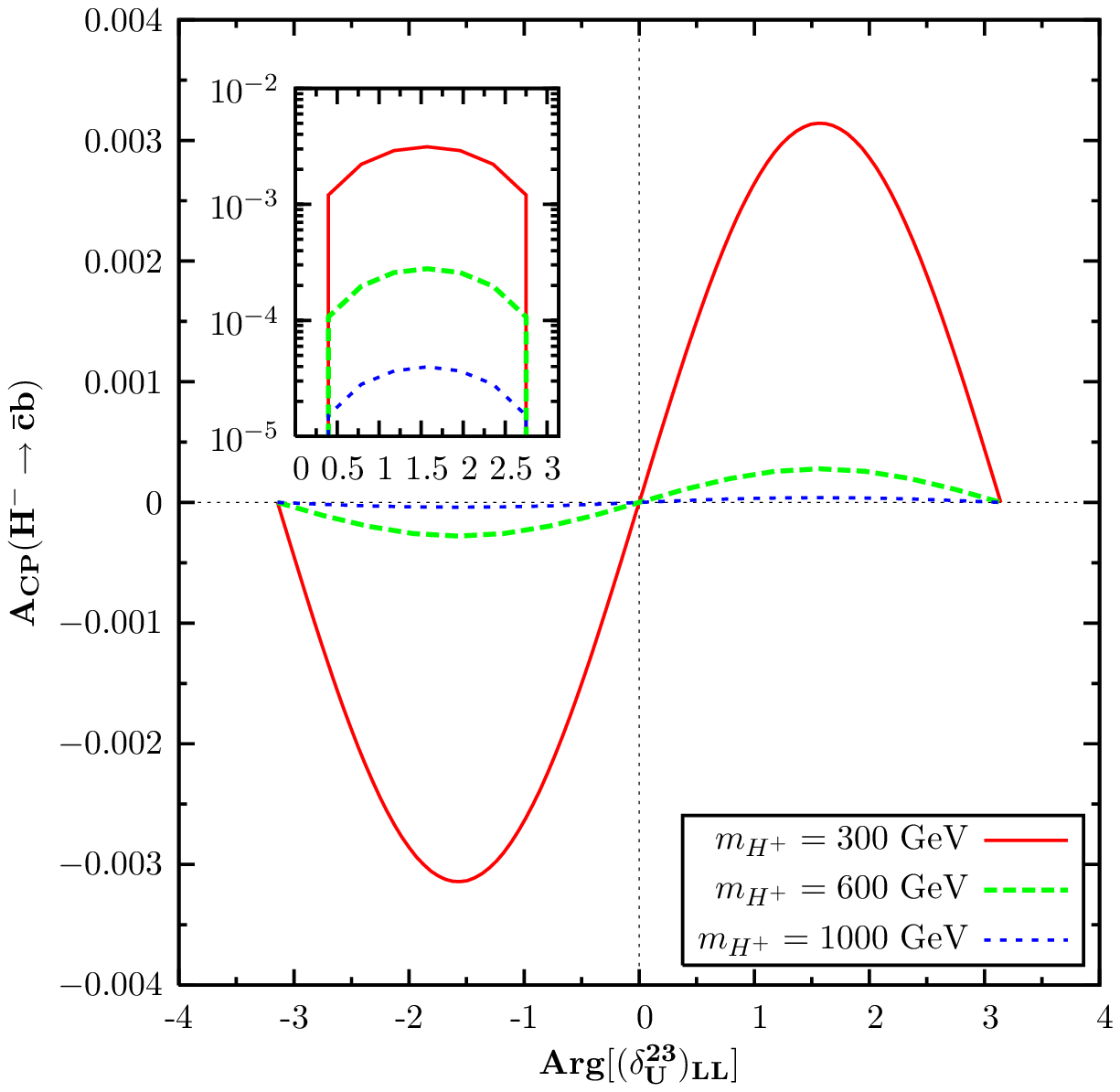} &
	\includegraphics[width=3.24in]{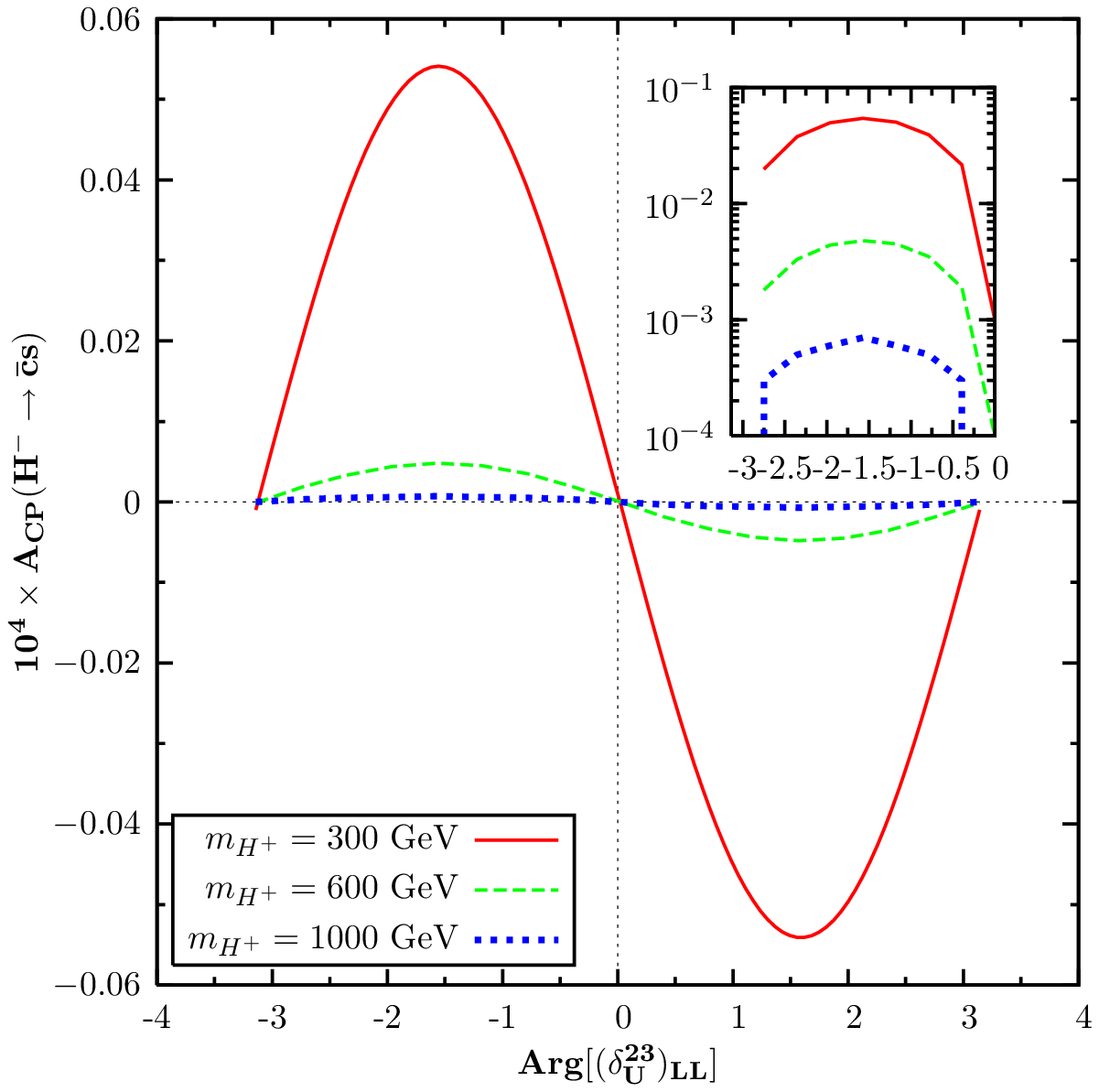} \\
\hspace*{-0.5cm}
	\includegraphics[width=3.22in]{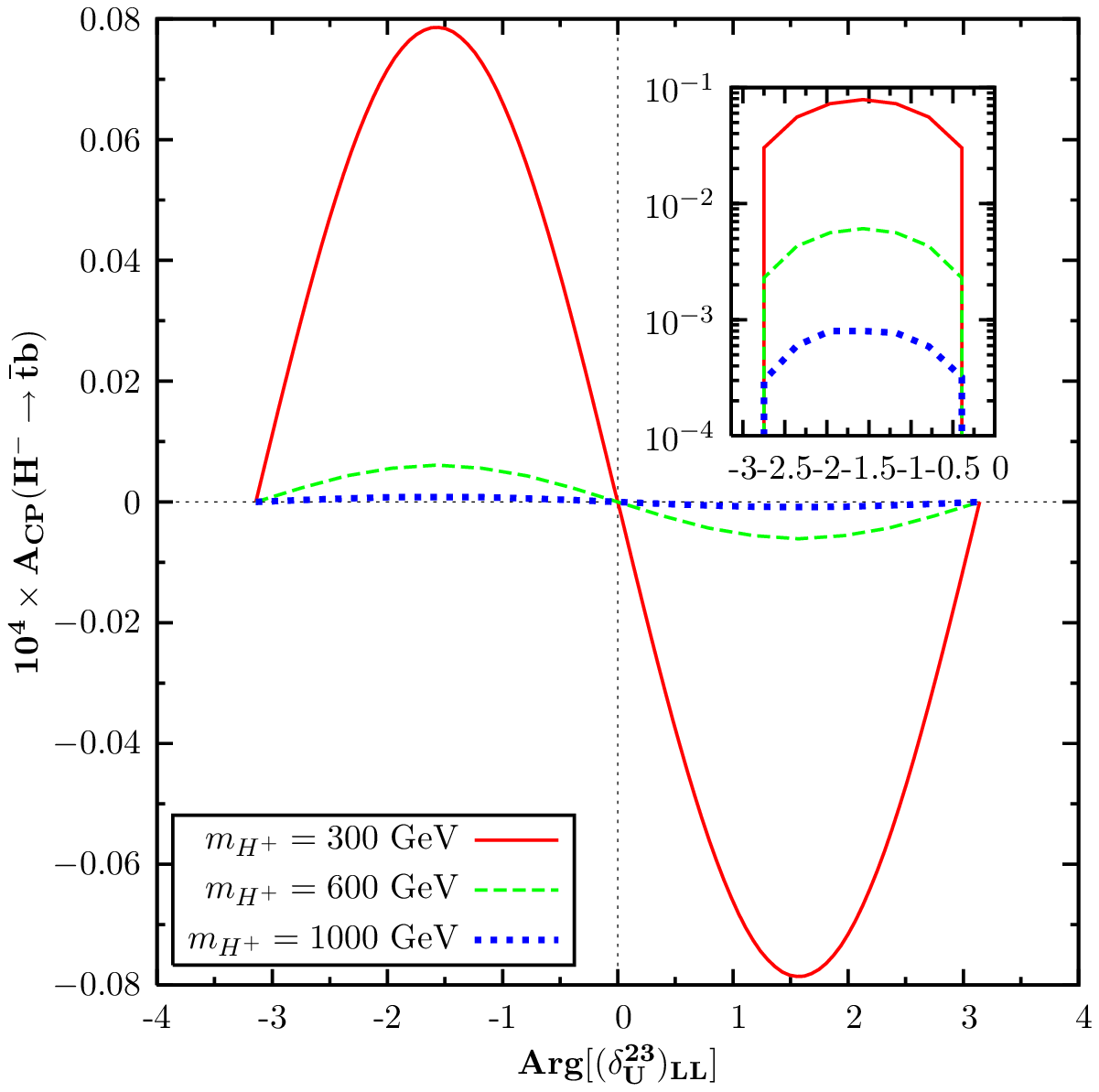} &
	\includegraphics[width=3.23in]{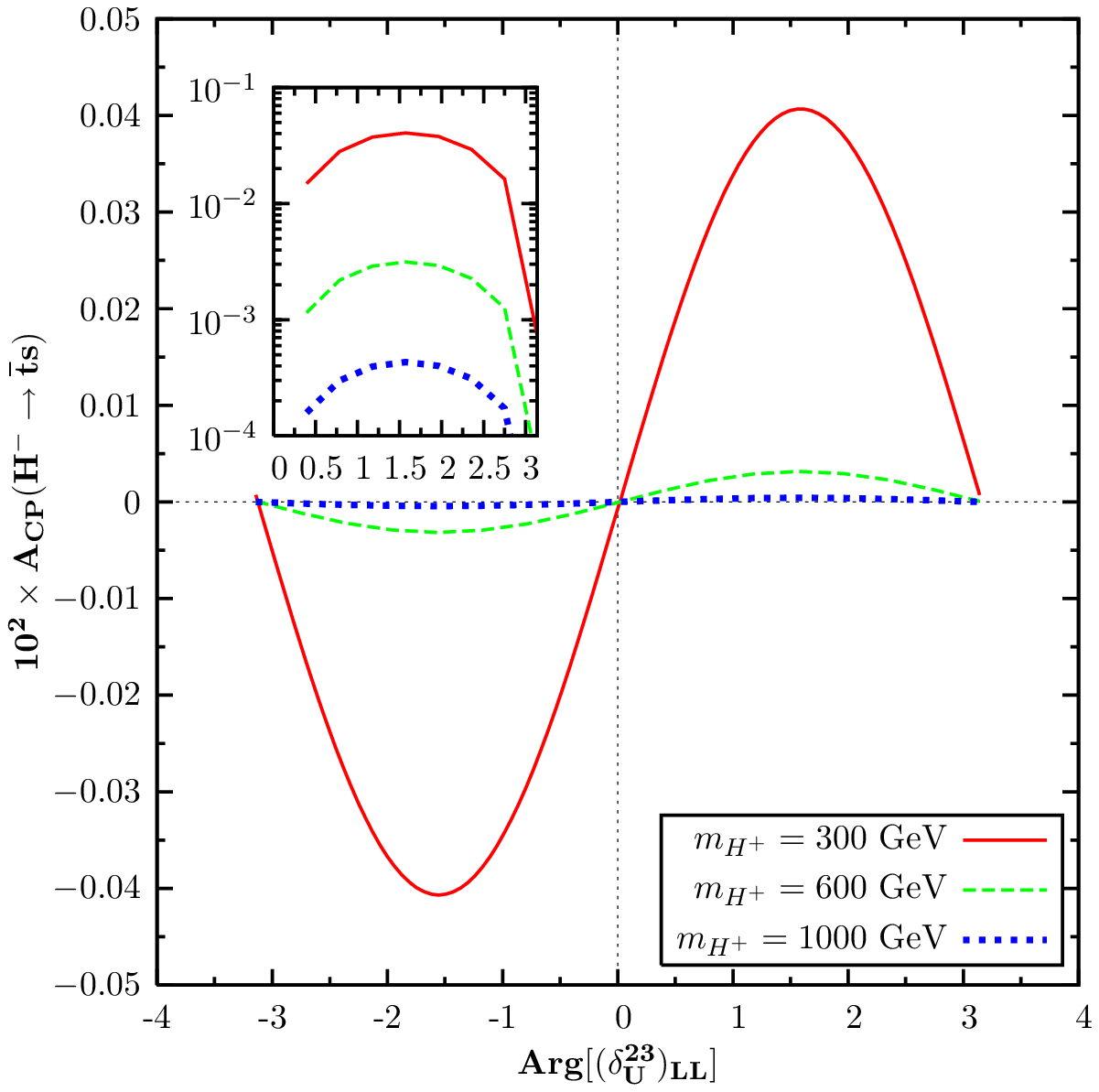}
	\end{array}$
\end{center}
\vskip -0.2in
      \caption{The same as Fig.~\ref{fig:phiA} but as a function of the phase of the flavor violating parameter ${\rm Arg}[(\delta_U^{23})_{LL}]$ for various $m_{H^+}$ values. We take $A=1200$ GeV and ${\rm Arg}[A]=0$. The absolute value of $(\delta_U^{23})_{LL}$ is set to 0.5. The small graphs inside each graph represent the positive asymmetry $A_{CP}$ in the logarithmic scale.}
\label{fig:phiULL}
\end{figure}
\begin{figure}[htb]
\begin{center}$
	\begin{array}{cc}
\hspace*{-0.5cm}
	\includegraphics[width=3.2in]{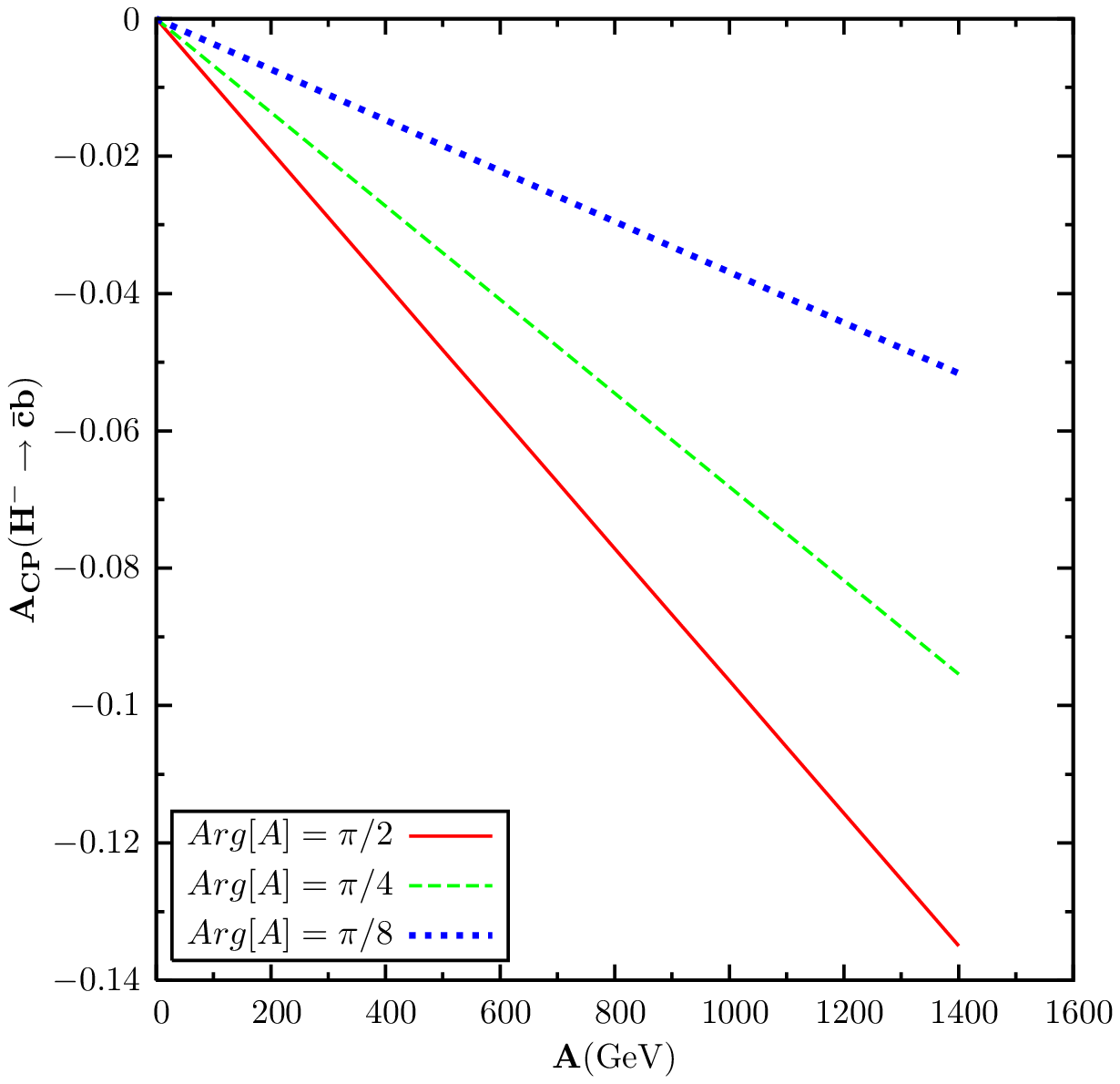} &
	\includegraphics[width=3.2in]{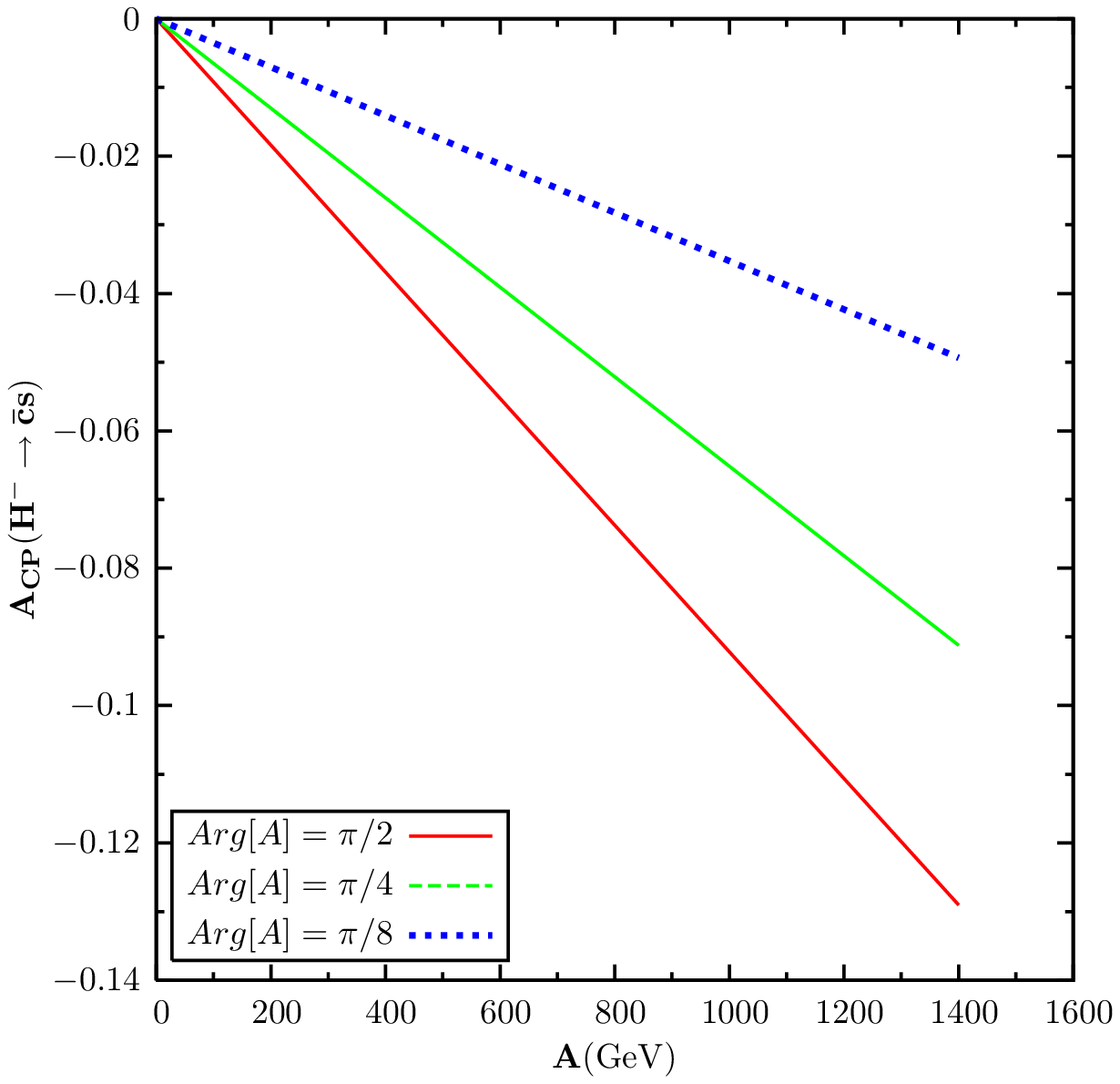} \\
\hspace*{-0.3cm}
	\includegraphics[width=3.2in]{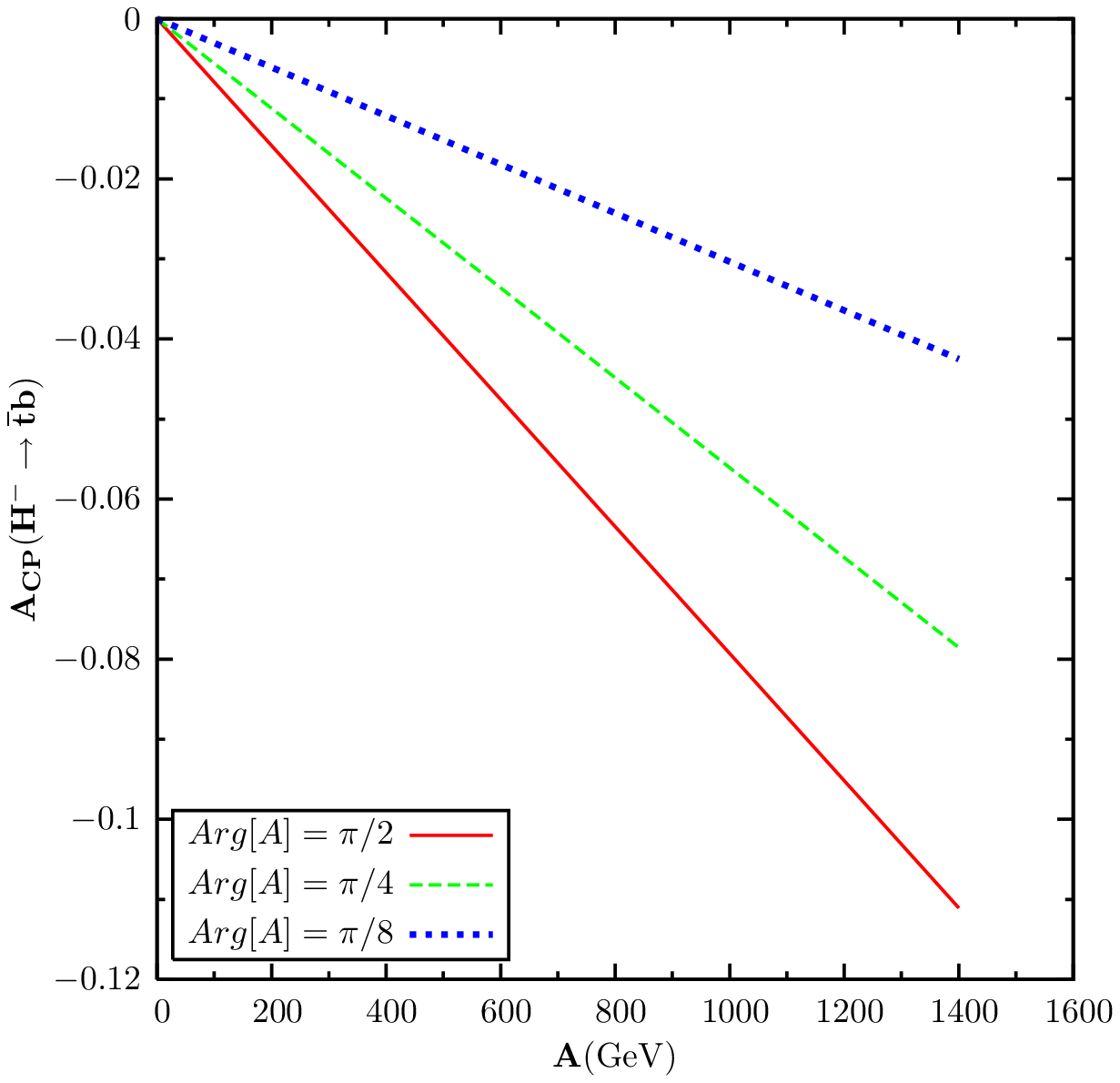} &
	\includegraphics[width=3.2in]{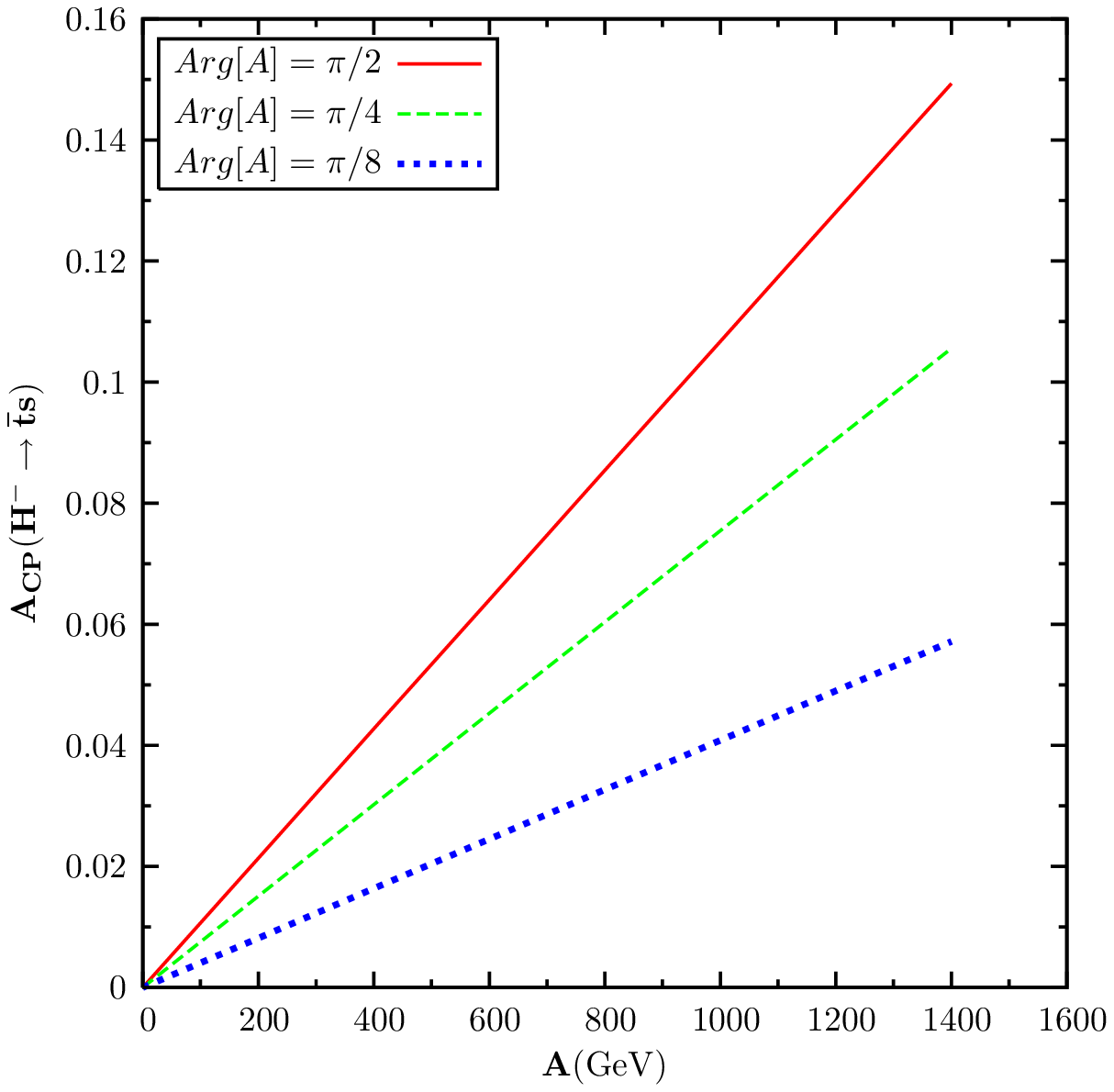}
	\end{array}$
\end{center}
\vskip -0.2in
      \caption{The $CP$ asymmetry for the charged Higgs decays $H^-\to \bar{u}_i d_j$ as a function of $A$ for the various values of the phase ${\rm Arg}[A]$.}
\label{fig:A}
\end{figure}

Additionally, we want to test whether it is possible to account a sizable asymmetry in charged Higgs decays by introducing complex flavor mixing parameters between second and third generation quarks and keeping all the other parameters real. This is  a secondary effect since, unlike the case with no-zero phase ${\rm Arg}[A]$, getting the absorptive phase through such parameters requires not only a chirality flip in the squark propagators but also a mass insertion for flavor changing as well (this is a CP breaking and flavor violating effect). In Fig.~\ref{fig:phiULL} we show the asymmetry 
$A_{CP}$  as a function of the phase ${\rm Arg}[(\delta_U^{23})_{LL}]$ for various charged Higgs masses. The small graphs inside each graph represent the positive asymmetry $A_{CP}$ in the logarithmic scale, so that one can distinguish the curves with  different charged Higgs masses. With the exception of the $H^-\to\bar{c}b$ decay mode, which can have as large as $0.3\%$ CP asymmetry for $m_{H^+}=300\,$ GeV, the other decays yield negligible asymmetries. The absolute value of $(\delta_U^{23})_{LL}$ is set to 0.5, but $A_{CP}$ is not very sensitive to the value of this parameter. We also checked the asymmetry induced by the phases of the other $\delta$ parameters in both the up and down sectors, but they all give negligible contributions. Therefore, it is unlikely that a measurable CP asymmetry in charged Higgs decays $H^-\to \bar{u}_i d_j$ is generated only by absorptive phase from $(\delta_U^{23})_{LL}$. 
\begin{figure}[htb]
\begin{center}$
	\begin{array}{cc}
\hspace*{-0.5cm}
	\includegraphics[width=3.1in]{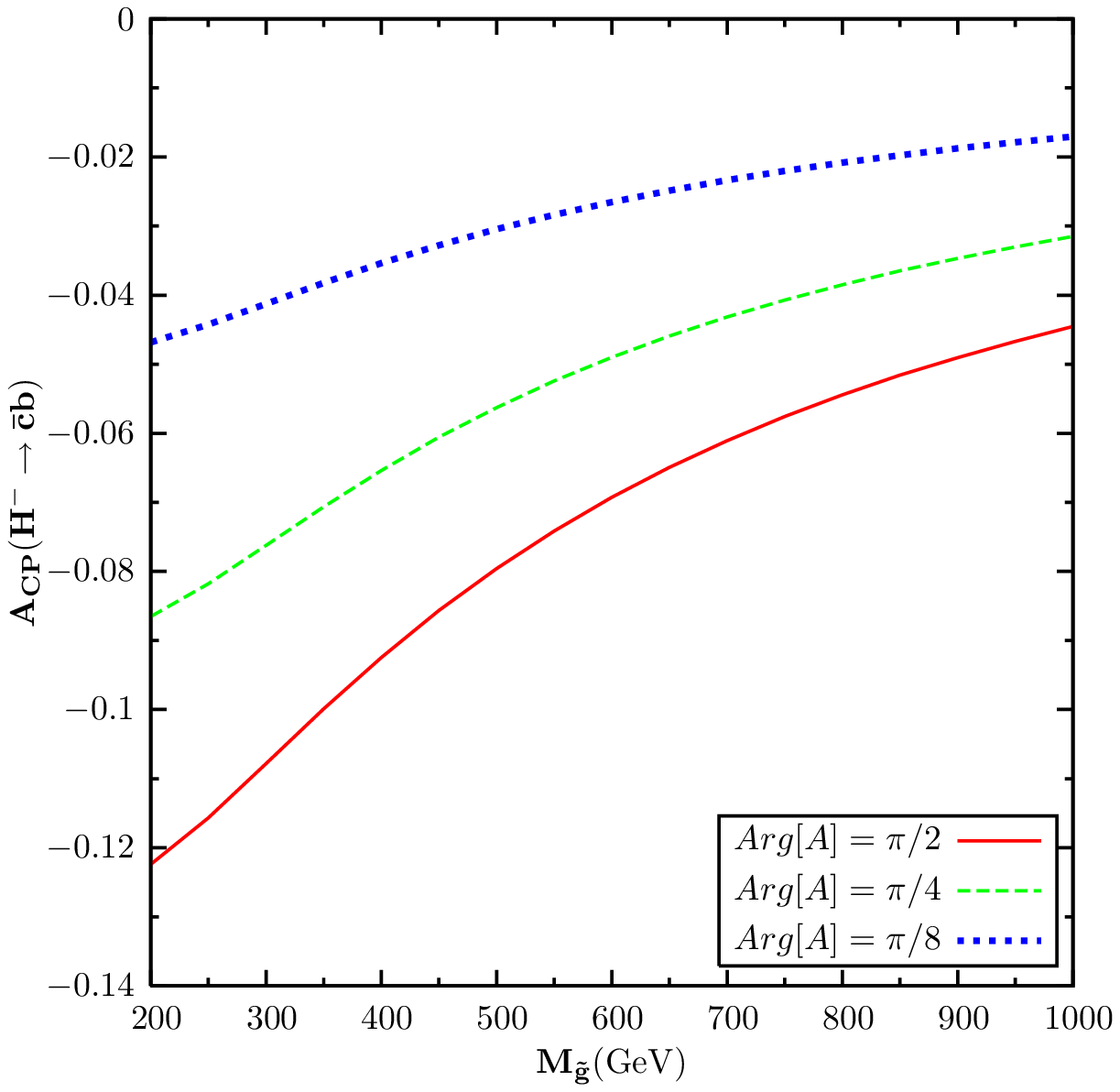} &
	\includegraphics[width=3.1in]{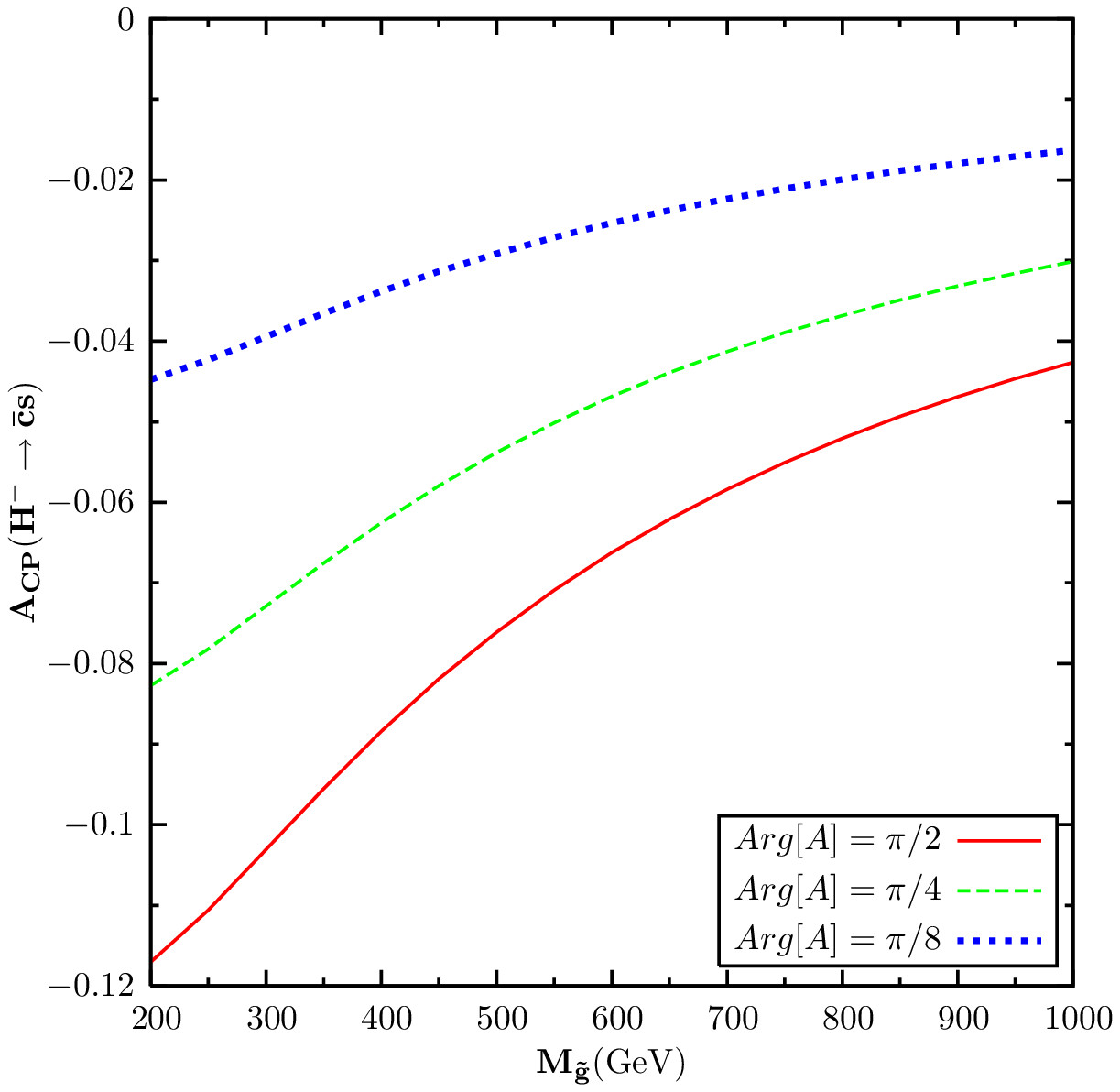} \\
\hspace*{-0.5cm}
	\includegraphics[width=3.15in]{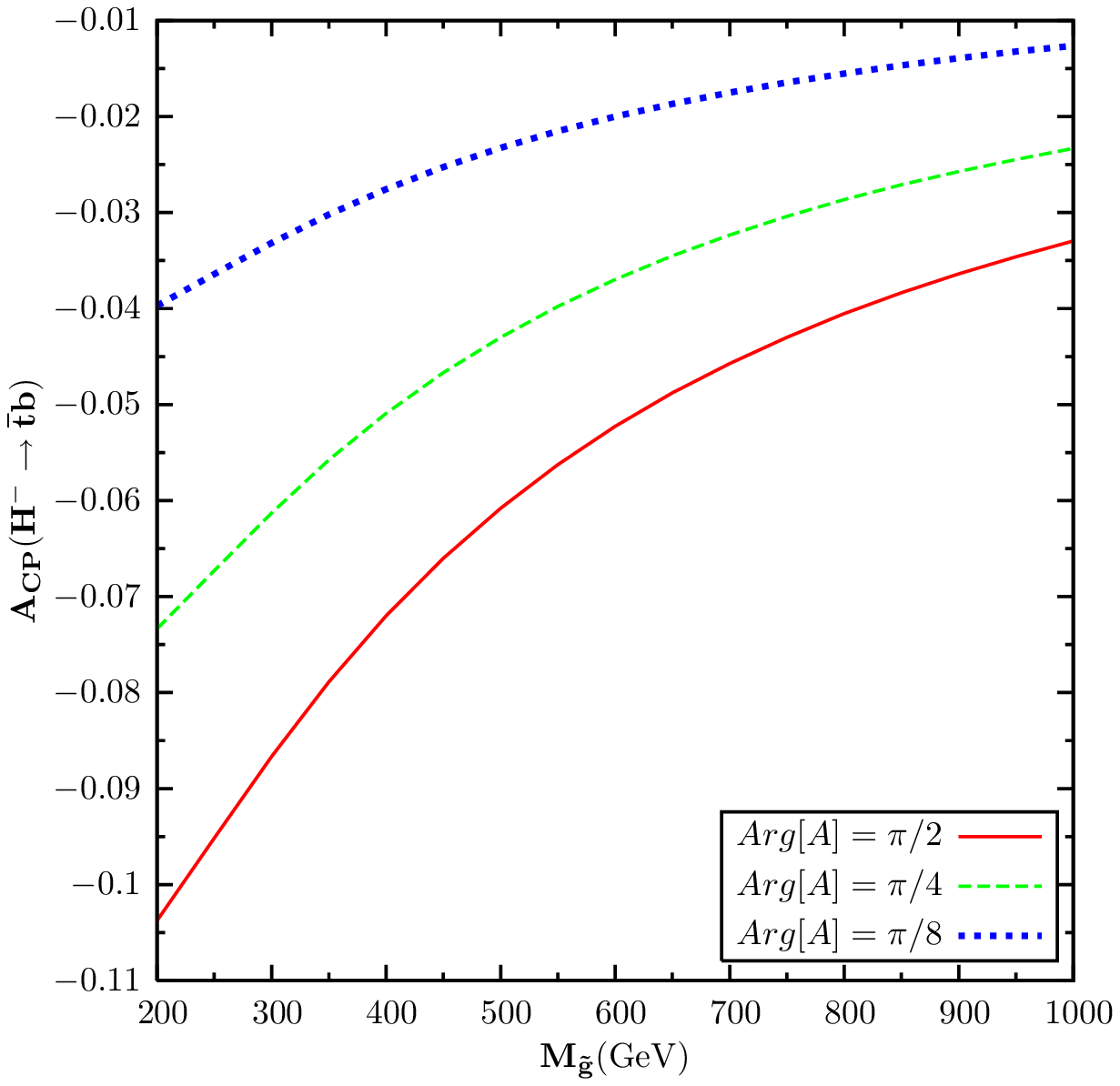} &
	\includegraphics[width=3.1in]{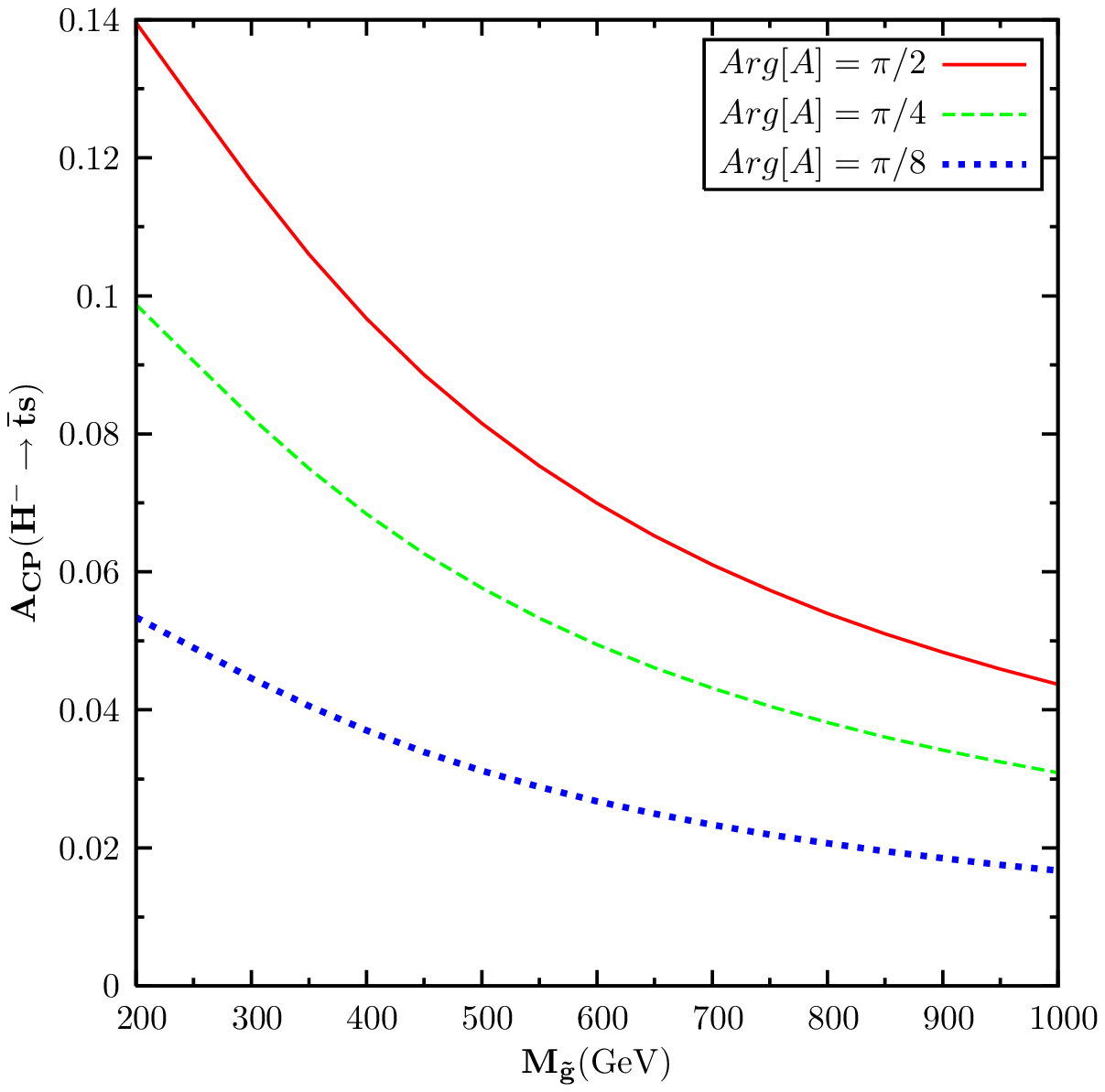}
	\end{array}$
\end{center}
\vskip -0.2in
      \caption{The same as Fig.~\ref{fig:A} but as a function of the gluino mass $m_{\tilde{g}}$. $A=1200$ GeV is assumed.}
\label{fig:MGl}
\end{figure}
\begin{figure}[htb]
\begin{center}$
	\begin{array}{cc}
\hspace*{-0.5cm}
	\includegraphics[width=3.1in]{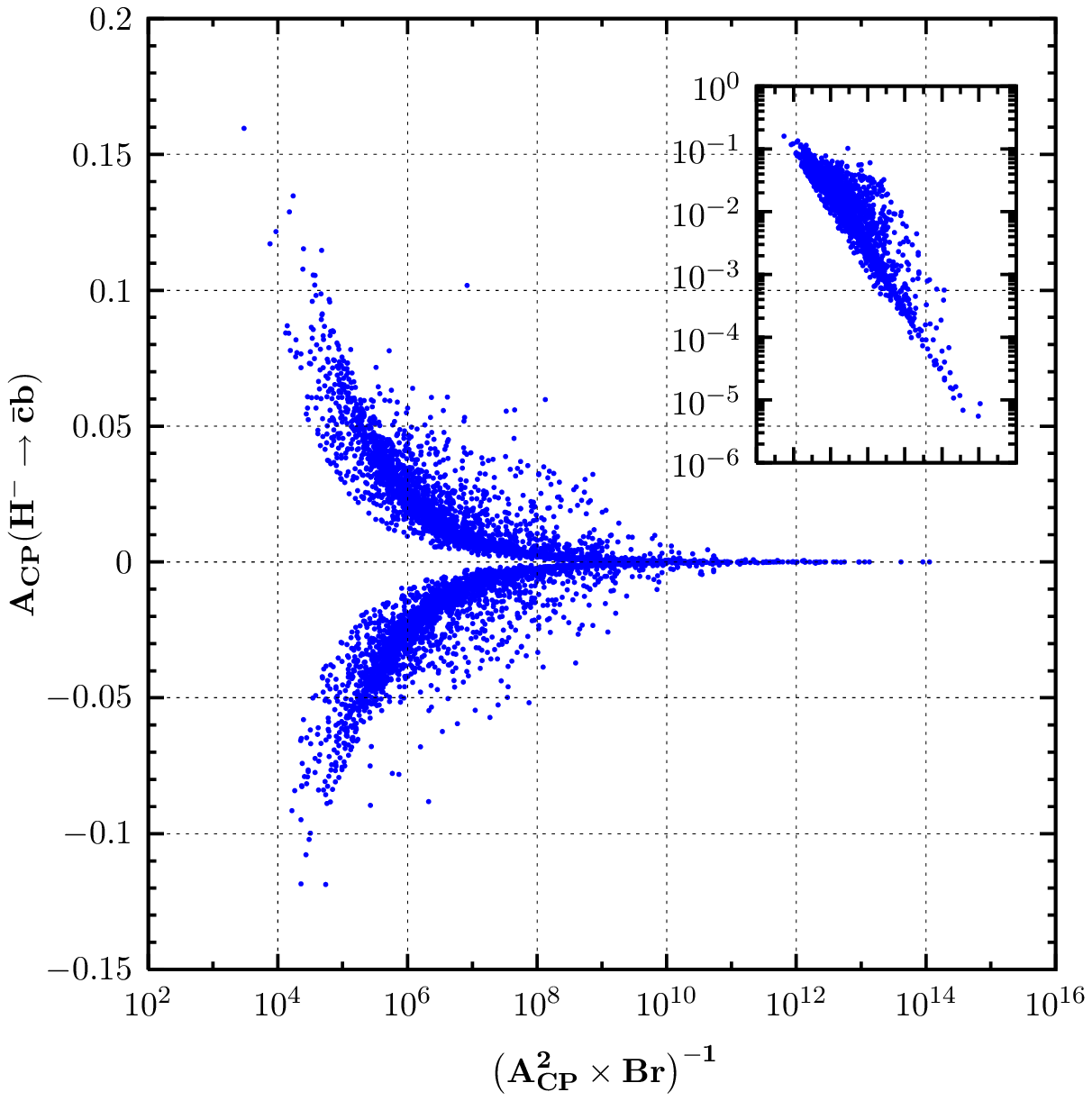} &
	\includegraphics[width=3.13in]{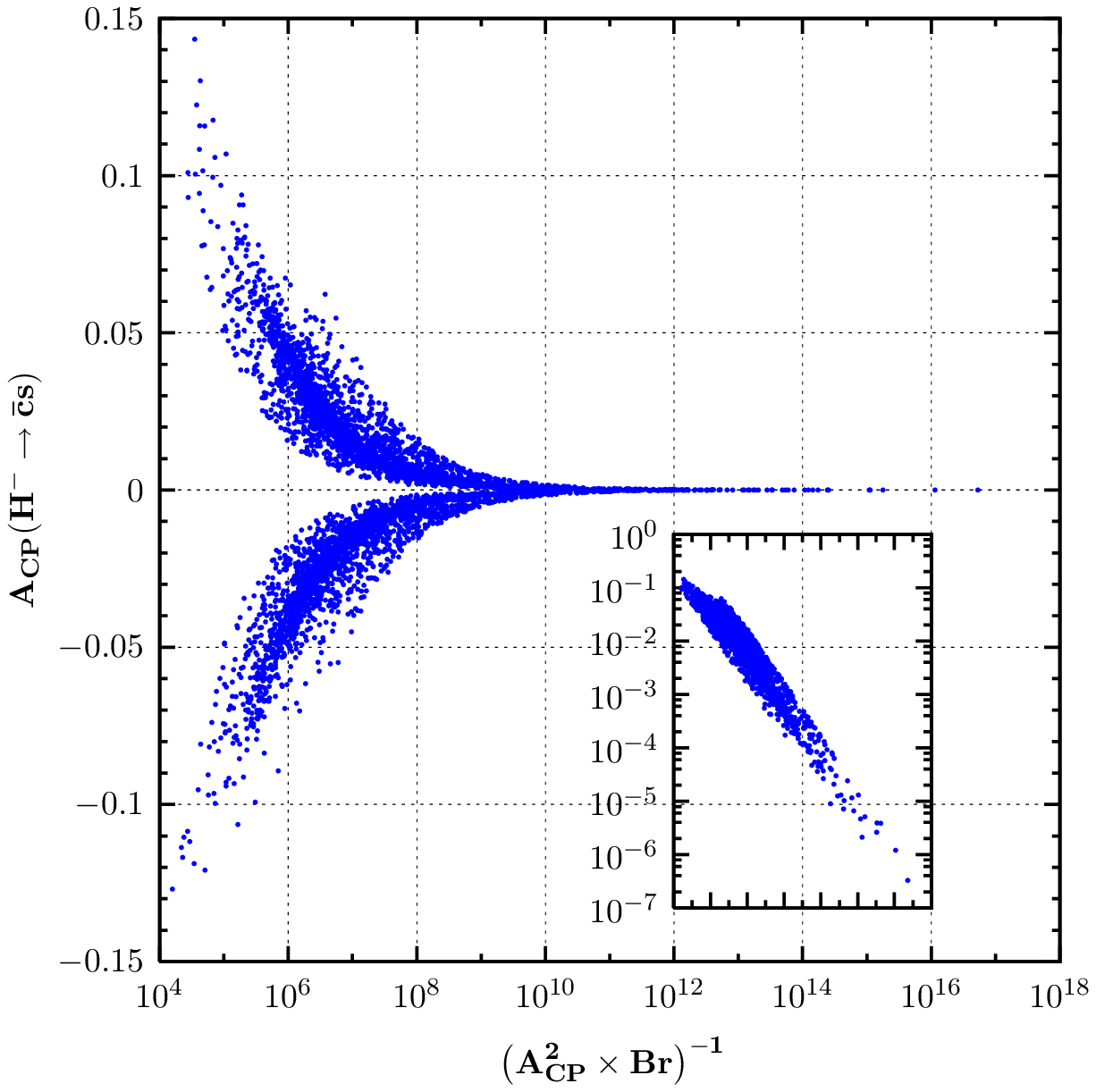} \\
\hspace*{-0.5cm}
	\includegraphics[width=3.1in]{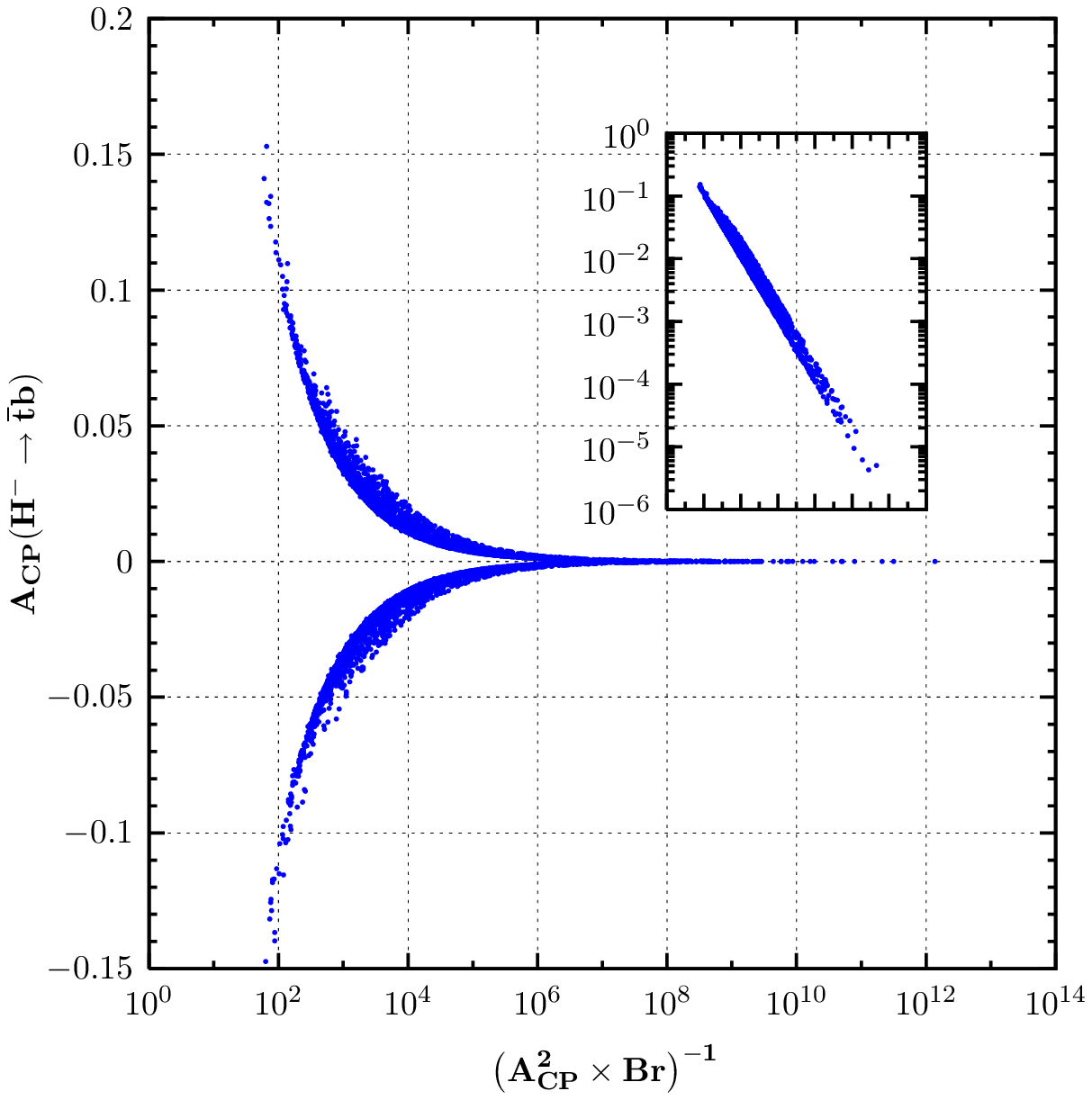} &
	\includegraphics[width=3.1in]{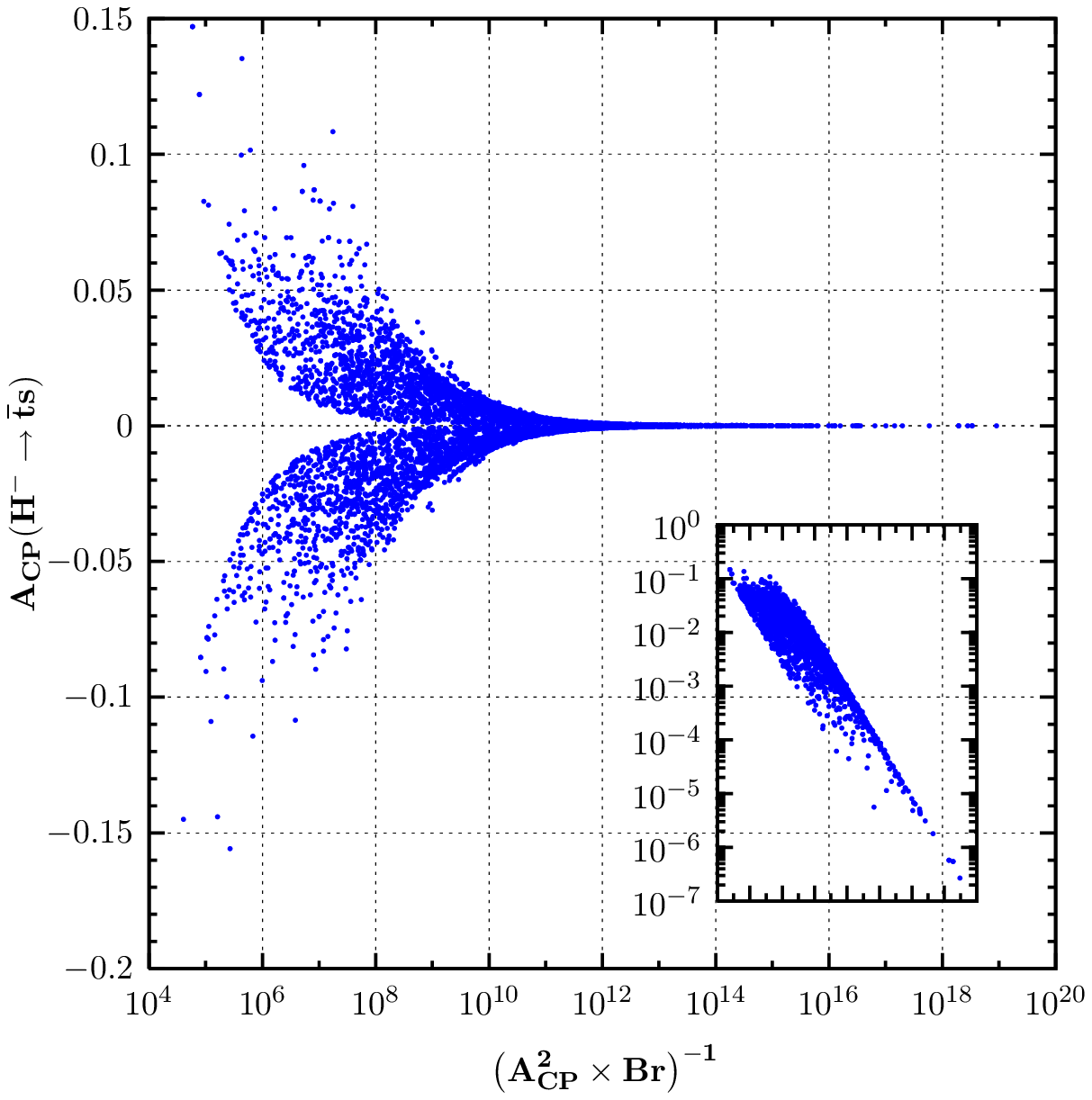}
	\end{array}$
\end{center}
\vskip -0.2in
      \caption{The scatter plot for the charged Higgs decays $H^-\to \bar{u}_i d_j$ in the $(A_{CP}^2\times Br)^{-1}-A_{CP}$ plane. These are obtained by scanning the sensitive parameters $A\in(0,1400)\,{\rm GeV},\; {\rm  Arg}[A]\in(-\pi,\pi),\; m_{\tilde g}\in(200,1000)\,{\rm GeV},$ $m_{H^+}\in(200,1000)\,{\rm GeV}$, and $\tan\beta\in(1,50)$. The x-axes are in the logarithmic scale. The small graphs inside each graph represent the positive asymmetry $A_{CP}$ in the logarithmic scale.} \label{fig:scatter1}
\end{figure}

In the next two figures, Figs.~\ref{fig:A} and \ref{fig:MGl}, we plot the CP asymmetry $A_{CP}$, for various ${\rm Arg}[A]$ values, as a function of $A$ and the gluino mass $m_{\tilde g}$, respectively. $A_{CP}$ in both of the decays $H^-\to \bar{c}b$ and $H^-\to\bar{c}s$  in Fig.~\ref{fig:A} can be as large as $14\%$ at ${\rm Arg}[A]=\pi/2$ for large $A$ values, but $H^-\to \bar{t}b$ remains slightly smaller. The CP asymmetry in $H^-\to \bar{t}s$ becomes even bigger, but  this has to be taken with some care. More than $10\%$ asymmetry could be achieved for each decay mode for $A=1200\, {\rm GeV}$ and if the gluino is very light ($\sim 200\, {\rm GeV}$), as shown in Fig.~\ref{fig:MGl}.  We note that there exists similar dependence on the charged Higgs mass.  The CP asymmetry also changes between $\pm15\%$ as we vary $\tan\beta$. The dependence is significant especially for light charged Higgs masses and small $\tan\beta$ values.
\begin{figure}[htb]
\begin{center}$
	\begin{array}{cc}
\hspace*{-0.5cm}
	\includegraphics[ width=3.1in]{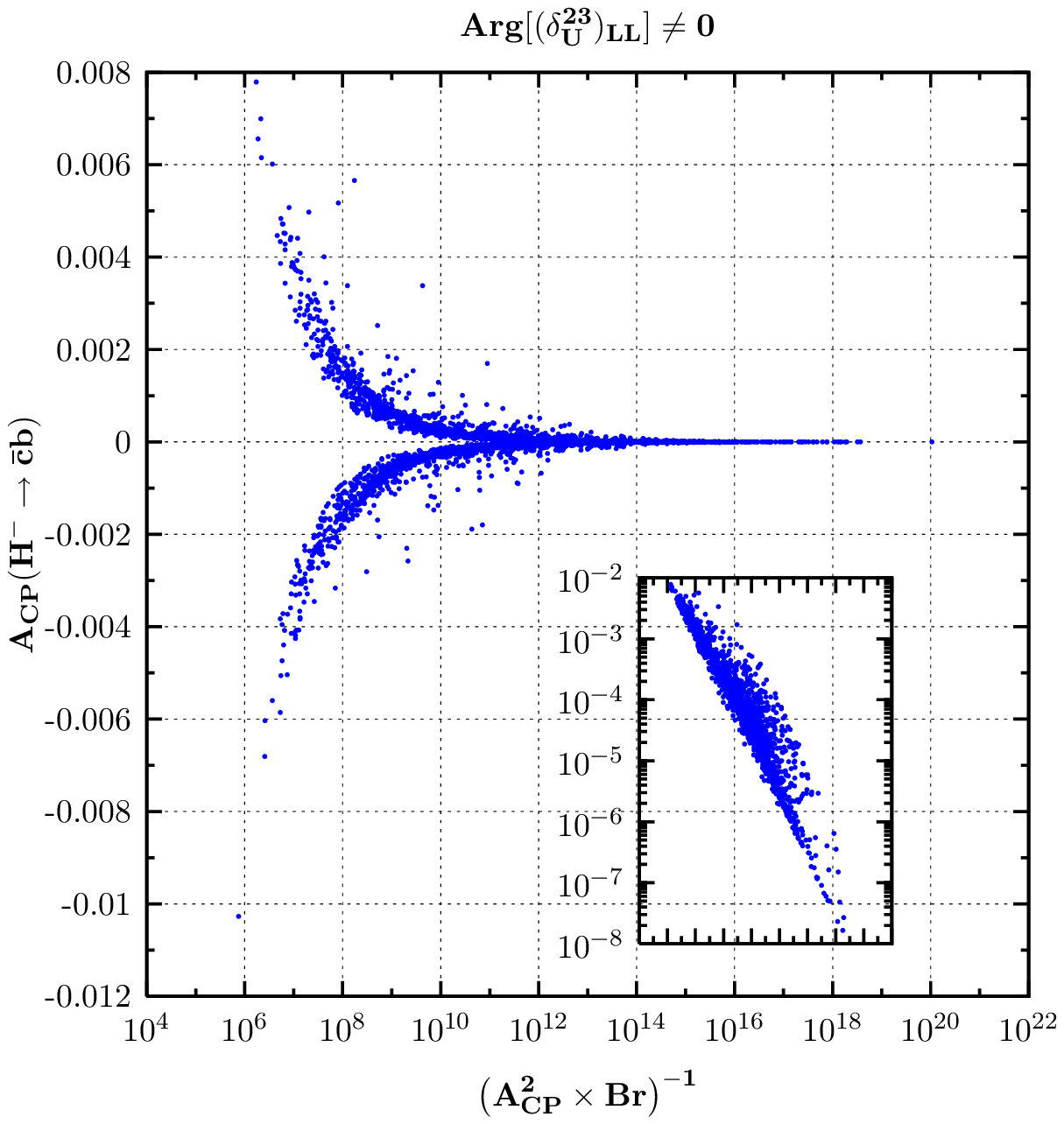} &\hspace*{-0.2cm}
	\includegraphics[width=3.2in]{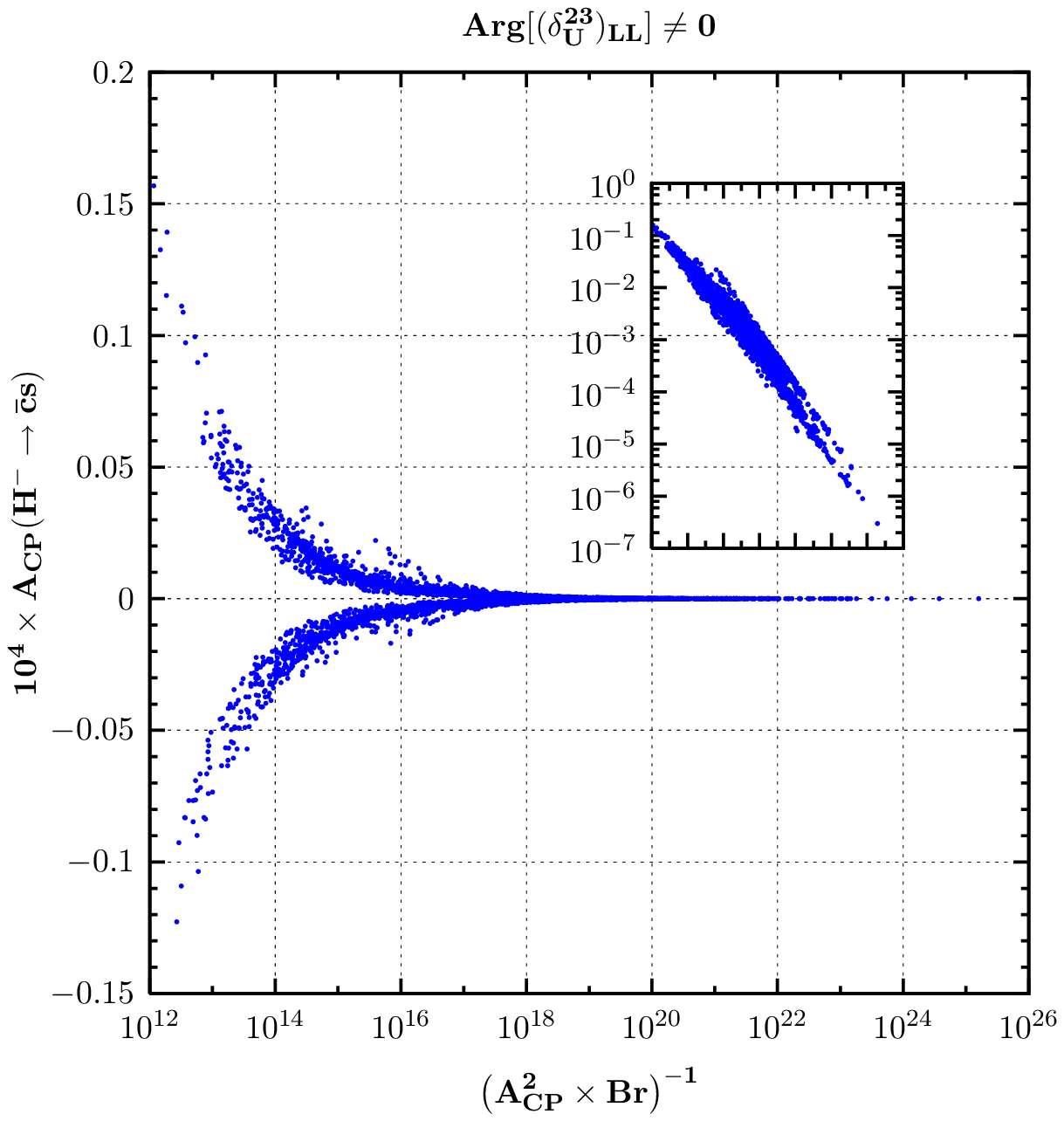} \\
\hspace*{-0.8cm}
	\includegraphics[width=3.26in]{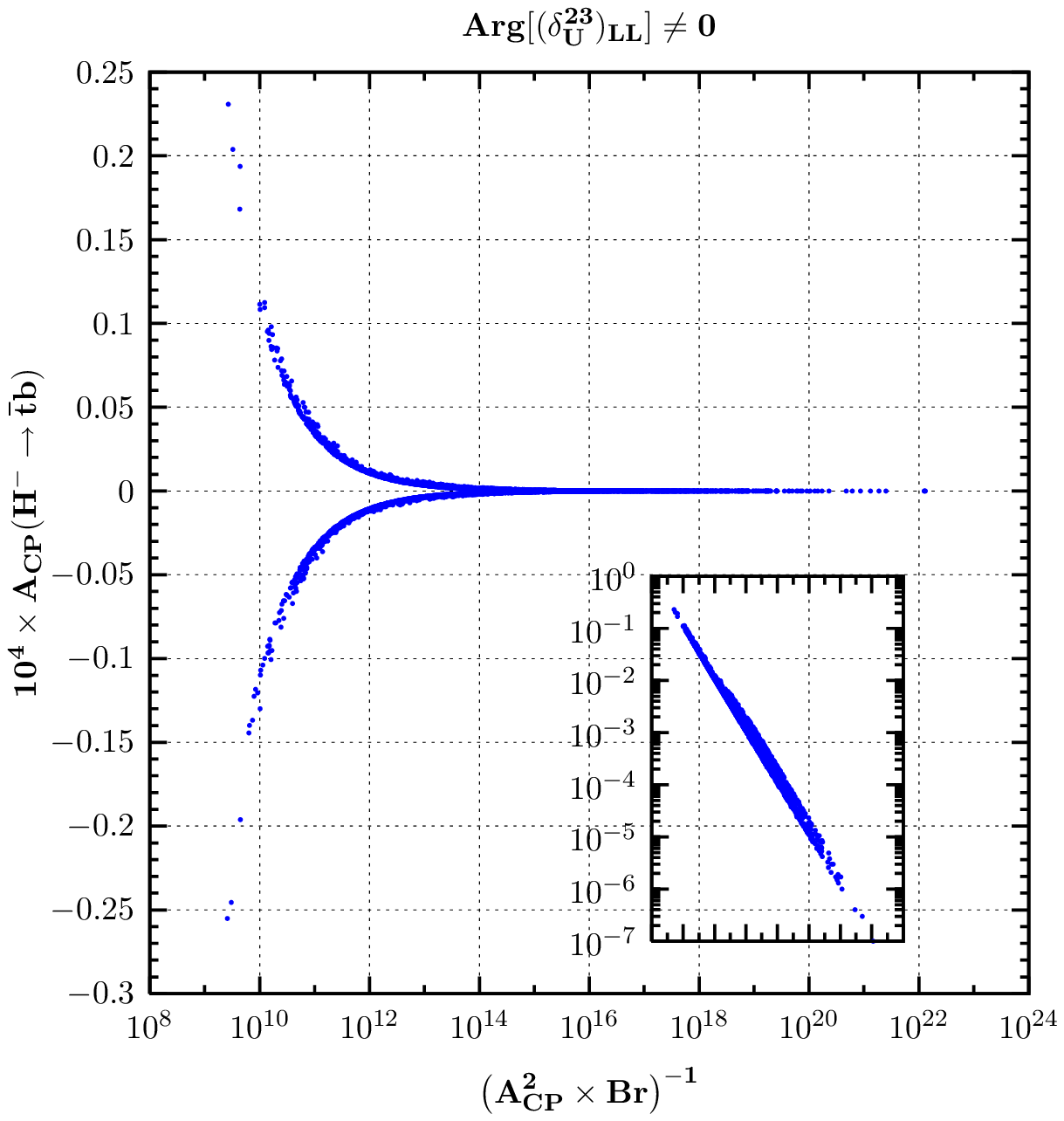} &
	\includegraphics[width=3.13in]{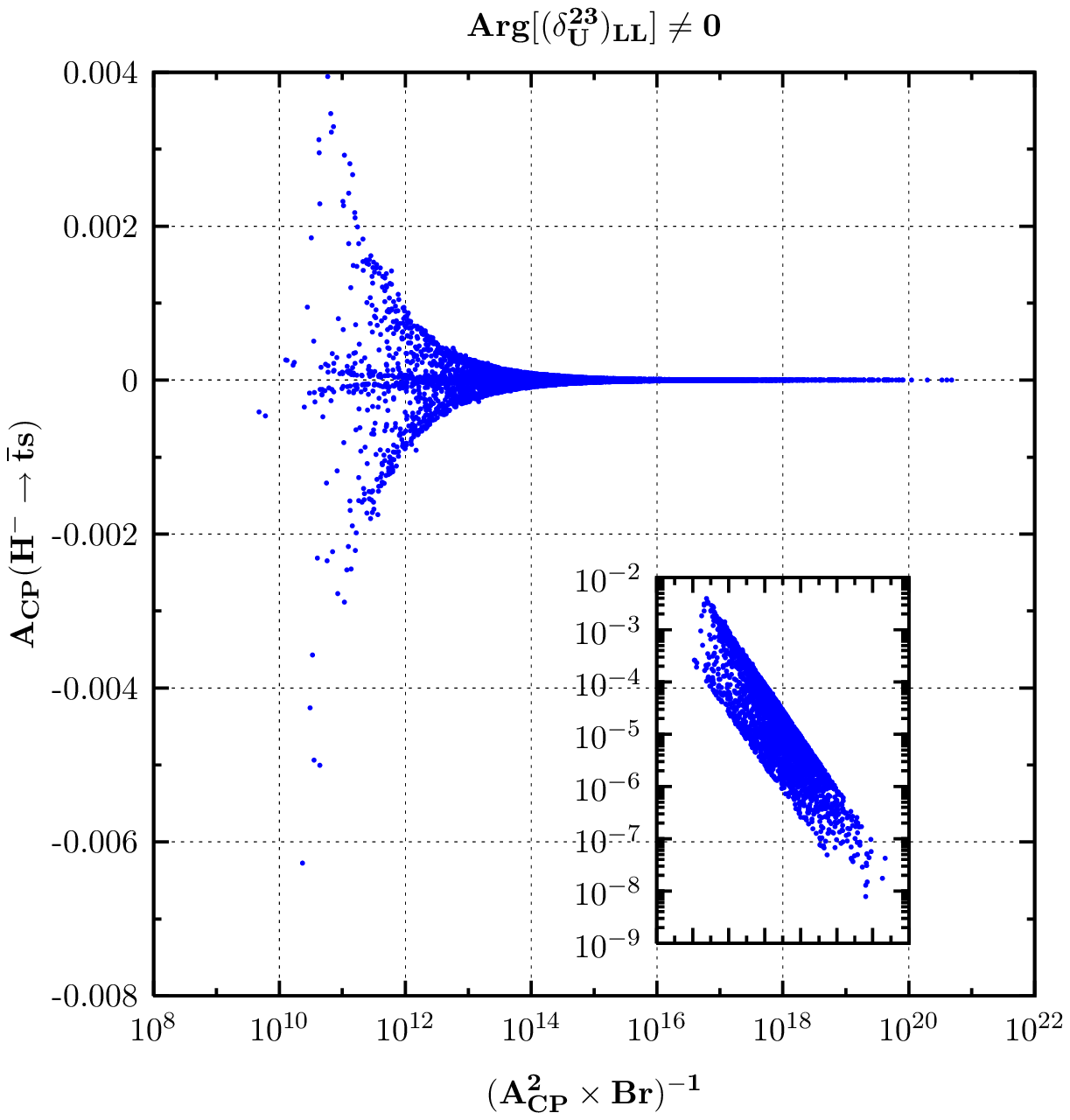}
	\end{array}$
\end{center}
\vskip -0.2in
      \caption{The same as Fig.~\ref{fig:scatter1} but for non-zero ${\rm Arg}[(\delta_U^{23})_{LL}]$ phase. $A$ is set to $1200\, {\rm GeV}$ and taken real. The x-axes are in the logarithmic scale. The small graphs inside each graph represent the positive asymmetry $A_{CP}$ in the logarithmic scale.}
\label{fig:scatter2}
\end{figure}

In order to test the viability of these channels for the search of CP asymmetry, we scanned the parameter space spanned by the most sensitive decay channels and compared the number of events (which can taken as the required number of charged Higgs bosons) as function of $(A_{CP}^2\times Br)^{-1}$.  In Fig.~\ref{fig:scatter1}, we show the scatter plots for the charged Higgs decays $H^-\to \bar{u}_i d_j$ in the $(A_{CP}^2\times Br)^{-1}-A_{CP}$ plane. These events are obtained by running the sensitive parameters randomly in the parameter ranges $A\in(0,1400)\,{\rm GeV},\; {\rm Arg}[A]\in(-\pi,\pi),\; m_{\tilde g}\in(200,1000)\,$GeV,  $m_{H^+}\in(200,1000)\,$GeV, and $\tan\beta \in(1,50)$. Note that the x-axes are in logarithmic scale. The branching ratios are calculated by evaluating all the other tree level charged Higgs decay widths  with  the  {\tt FeynHiggs} program.  Again, the small graphs show the positive part of the asymmetry  distribution in the logarithmic scale. From these scatter plots, one can make a few observations. 

Each decay channel has a possibility to develop an asymmetry bigger than $10\%$. In the case of $H^-\to\bar{c}s$ and $H^-\to\bar{t}s$, there are quite number of events which yield asymmetries around $15\%$, which is as large an asymmetry as the other two channels can attain. But the number of charged Higgs bosons required to observe such an asymmetry is around $10^5-10^6$ for $\bar{t}s$ decay, which is $2$ to $4$ orders of magnitude more than the number needed to make such a measurement in the $H^-\to\bar{c}b$ or $H^-\to\bar{t}b$ channel, respectively. On the other hand, $H^-\to\bar{c}s$ yields a comparable asymmetry distribution with respect to $H^-\to\bar{c}b$, and also the required number of charged Higgs is similar for some part of the parameter space which maximizes the asymmetry. Clearly, the parameter set for the maximal scenario is different for each decay modes. Such a similarity between $\bar{c}s$ and $\bar{c}b$ channels is not surprising since  the $\bar{c}s$ decay mode can have a branching ratio  comparable to, or larger than, that of $\bar{c}b$ decay mode, depending on the value of $\tan\beta$. In the $\bar{c}s$ channel, since the strange quark is light, it is difficult to distinguish it from $\bar{c}d$ channel.

 So, we can conclude that,  depending mainly  on the $\tan\beta$ value, the asymmetry in both $\bar{c}b$ and $\bar{c}s$ can be competitive with the one from $\bar{t}b$ if there are enough charged Higgs bosons produced at the colliders. But one needs at least $10^2$ times more statistics for $\bar{c}b$ and $\bar{c}s$ cases. From Fig.~\ref{fig:scatter1}, the distribution of the events for $\bar{t}b$ modes, unlike the other channels,  are not scattered much since in almost the entire parameter range considered in the scan, its branching ratio remains constant.
 
In the last figure, Fig.~\ref{fig:scatter2}, we perform the same scan of  the parameter space as before, but with a non-zero phase  ${\rm Arg}[(\delta_U^{23})_{LL}]$ in the up-type squark mixing matrix, while switching the phase of $A$ off. So, instead of running the trilinear coupling $A$ and its phase, we run $(\delta_U^{23})_{LL}$ and its phase in the range $(0,0.5)$ and $(-\pi,\pi)$, respectively. It seems that the largest asymmetry comes from the $H^-\to\bar{c}b$ process which could be at most as large as $0.6-0.8\%$,  but at least $10^6$ number of Higgs produced are required to probe such asymmetry. Additionally, the asymmetry is still significantly smaller than the case where the absorptive phase originated from the trilinear couplings. For the other channels, not only  are the asymmetries very small, but as a consequence the number of charged Higgs required  to observe them is enormous. 

\section{Summary and Conclusion}
\label{conc}
In this study, we analyzed the possibility of obtaining measurable signals for CP asymmetry in the charged Higgs decays $H^-\to \bar{u}_i d_j$ for the second and third generation quarks in the MSSM. Above the top quark threshold, charged Higgs bosons decay mainly to $\bar{t}b$,  but decays to $\bar{c}b$, $\bar{c}s$, and $\bar{t}s$ are also relevant. The CP asymmetry of the main decay mode 
$\bar{t}b$ and other non-quark charged Higgs decays have been considered previously \cite{Christova:2002ke,Christova:2002sw,Christova:2006fb}. Here we analyzed,  discussed and compared all significant quark channels to pinpoint which one is more likely to produce a visible asymmetry at the colliders.  A non-zero CP asymmetry requires an absorptive phase, for which we considered possible interference terms between tree-level and one-loop diagrams. Then we calculated the imaginary parts of one-loop scalar two- and three-point diagrams by first deriving first generic analytical formulas for the discontinuity in such diagrams with the help of Cutkosky rules. We then presented the results obtained for the vectorial and tensorial types of integrals. Within the MSSM framework, we investigated the effect on the asymmetry of two relevant absorptive phases; the common phase of the trilinear couplings $A$, and the  phase from flavor the violating parameter $(\delta_U^{23})_{LL}$. We analyzed their effects separately and without considering any interference effects. 

Consideration of non-zero values for ${\rm Arg}[A]$ predicted asymmetries around $10-15\%$ for each decay mode. However, by including the requirement that $(A_{CP}^2\times Br)^{-1}$ is proportional to the number of required Higgs bosons, our analysis indicates that only $H^-\to \bar{t}b$ and $H^-\to \bar{c}b,\,\bar{c}s$ decay processes would be likely to induce a measurable CP asymmetry, with the requirement that at least $10^2-10^4$ charged Higgs bosons be produced at the colliders.  Unlike the phase ${\rm Arg}[A]$, the phase $(\delta_U^{23})_{LL}$ cannot account a sizable CP asymmetry $A_{CP}$. In this case, only $H^-\to\bar{c}b$ can get an asymmetry around $0.6-0.8\%$, with the requirement that $10^6$ charged Higgs bosons must be produced at the colliders, in order to translate into a measurable asymmetry. The fact that the arguments of $A$ and $(\delta_U^{23})_{LL}$ give rise to CP asymmetries of different orders of magnitude justifies {\it a posteriori} neglecting interference effects.

\section{Acknowledgment}
This work is supported in part by NSERC of Canada under the Grant No. SAP01105354.

\appendix*
\section{The Method - Calculation of the Imaginary Part}
We calculate the absorptive part of the loop diagrams by applying the Cutkosky rules. In general, in the loop integration, we end up with a numerator with scalar, vectorial, or tensorial structures (and their pseudo counterparts), depending on the types of particles running in the loop. We first consider the scalar case and then outline the method for the vectorial and tensorial cases and present the results. We discuss self energy and vertex cases separately.
\subsubsection{Two-Point Function:}
In Fig.~\ref{fig:cuts}-(a), if we assume that particles running in the loop are scalar, we get a $B_0$ type scalar two-point Passarino-Veltman function \cite{Passarino:1978jh}. We use the convention that all external momenta are incoming and the diagram (a) of Fig.~\ref{fig:cuts} can be expressed as
\begin{equation}
 B_0(p^2,m_1^2,m_2^2)=\frac{i}{\pi^2}\int d^4k \frac{1}{(k^2-m_1^2)\left((p-k)^2-m_2^2\right)}\;,
\end{equation}
where $m_1$ and $m_2$ are the masses of the particles in the loop. By applying the Cutkosky rules \cite{Cutkosky:1960sp} for Fig.~\ref{fig:cuts}-(a) we have, for the discontinuity across this cut,
\begin{eqnarray}
\Delta B_0(p_1^2,m_1^2,m_2^2)&=&\frac{i}{\pi^2}\int d^4k\,2\pi\,\Theta(k_1^0)\,\delta(k_1^2-m_1^2)\, 2\pi\,\Theta(p^0_1-k_1^0)\,\delta((p_1-k_1)^2-m_2^2)\nonumber\\
&=&2\pi i\, \frac{\sqrt{\lambda(p_1^2,m_1^2,m_2^2)}}{p_1^2}\,\Theta\left(p_1^2-(m_1+ m_2)^2\right),
\end{eqnarray}
where $k_1^0$ and $p_1^0$ are the energies of the corresponding particles. $\Delta B_0$ represent the discontinuity of $B^0$ and is related to the imaginary part of the diagram up to a factor of $2i$ (see \cite{Bauberger:1994hx} for example). $\lambda(x,y,z)$ is the usual K\"{a}llen function defined as $\lambda(x,y,z)=(x-y-z)^2-4yz$. The argument of the Heaviside function $\Theta$ in the final result ensures that both {\it internal-cut} states,  particles 1 and and 2, go on-shell in the loop. 

Now let's assume that we have a vectorial term in the numerator. If we denote this integral as $B^\mu$, then the discontinuity can be written
\begin{eqnarray}
\displaystyle
 \Delta B^\mu(p_1^2,m_1^2,m_2^2) &\equiv& \Delta B_0(p_1^2,m_1^2,m_2^2)\otimes k^\mu\nonumber\\
 &=&\displaystyle \left\{\begin{array}{r@{\quad,\quad}l}
            \frac{p_1^2+m_1^2-m_2^2}{2\,\sqrt{p_1^2}}\;\Delta B_0(p_1^2,m_1^2,m_2^2) & \mu =0 \\
            0 & \mu=i
            
           \end{array}\right.
\end{eqnarray}
where ``$\otimes k^\mu$'' means that $k^\mu$ should be considered as a part of the integrand of $\Delta B_0$ term.

In a similar manner, one can calculate a tensorial type of two-point integral. Since we will  need terms only up to second rank tensors in our calculation, $\Delta B^{\mu\nu}$ is enough to calculate for the discontinuity in such cases, and the result can be expressed as
\begin{eqnarray}
 \Delta B^{\mu\nu}(p_1^2,m_1^2,m_2^2) &\equiv& \Delta B_0(p_1^2,m_1^2,m_2^2) \otimes (k^\mu k^\nu)\nonumber\\
 &=&\kappa^{\mu\nu}\Delta B_0(p_1^2,m_1^2,m_2^2),
\end{eqnarray}
where 
\begin{eqnarray}
 \displaystyle
 \kappa^{\mu\nu} = {\rm Diag}(\kappa^0,\frac{\kappa}{3},\frac{\kappa}{3},\frac{\kappa}{3}),\;\;
 \kappa^0 = \frac{(p_1^2+m_1^2-m_2^2)^2}{4\,p_1^2},\;\;
 \kappa = \frac{\lambda(p_1^2,m_1^2,m_2^2)}{4\,p_1^2}\,.
\end{eqnarray}
Then the imaginary part of the self energy diagrams shown in  the last two diagrams of Fig.~\ref{fig:relevant} becomes the sum of the above terms
\begin{equation}
 {\mathcal Im}({\rm Self\; Energy}) = \frac{1}{2i}\,\sum_l\left( Y_l\,\Delta B_0 + Y_l^{\mu}\,\Delta B_{\mu} + Y_l^{\mu\nu}\,\Delta B_{\mu\nu}\right),
 \label{Yn}
\end{equation}
where $Y_l$'s include all other contributions arising from the Feynman rules and the index $l$ runs over loop diagrams. 
\subsubsection{Three-Point Function:}
 The evaluation is similar to the two-point function case but the calculation is more cumbersome. We first give the result for the discontinuity in a three-point scalar integral, known as $C_0(p_1^2,p_2^2,p_3^2,m_1^2,m_2^2,m_3^2)$\footnote{For simplicity, we suppress the argument of $C_0$ in the rest of the paper.}. Here $m_{1,2,3}$ are the internal masses. Again using Cutkosky rules, we have\footnote{Our result is consistent with the one given in \cite{Bauberger:1994hx}.} for the discontinuity
 \begin{eqnarray}
 \displaystyle
  \Delta C_0 &=& \frac{i}{\pi^2}\int d^4k_1 \frac{
  \,2\pi\,\Theta(k_1^0)\,\delta(k_1^2-m_1^2)\, 2\pi\,\Theta(-p_2^0-p_3^0-k_1^0)\,\delta\left((p_2+p_3+k_1)^2-m_2^2\right)}{(p_3+k_1)^2-m_3^2}\nonumber\\
  &=& \frac{-2\pi i}{\sqrt{\lambda(p_1^2,p_2^2,p_3^2)}}\; \log\left(\frac{\alpha+\beta}{\alpha+\beta}\right)\;\Theta(p_1^2-(m_1+m_2)^2),
 \end{eqnarray}
where
\begin{eqnarray}
 \alpha &=& p_1^2\left(p_1^2+2 m_3^2-(p_2^2+p_3^2+m_1^2+m_2^2)\right)-(m_1^2-m_2^2)(p_2^2-p_3^2)\,,\nonumber\\
 \beta  &=& \sqrt{\lambda(p_1^2,m_1^2,m_2^2)\;\lambda(p_1^2,p_2^2,p_3^2)}\,.
\end{eqnarray}

The vectorial type integrals can be calculated as follows. The extra term $k_1^\mu$ can be converted into the external momenta $p_i$'s and their derivatives with the help of the propagator in the denominator \cite{Scharf:1996zi}. For example, 
\begin{eqnarray}
 \frac{k_1^\mu}{(p_3+k_1)^2-m_3^2} =\frac{1}{2}\,\frac{\partial}{\partial p_{3\mu}}\left[\log\left((p_3+k_1)^2-m_3^2\right)\right]-\frac{p_3^\mu}{(p_3+k_1)^2-m_3^2}.
\end{eqnarray}
For convenience, we define $(p_3+k_1)^2-m_3^2\equiv D$, $f_0\equiv 1/D$, $f_1\equiv \log(D)$, and $f_2\equiv D\,(\log(D)-1)$. This way we can express the discontinuities in both the vectorial and the  $2^{nd}$ rank tensorial type integrals in the following form
 \begin{eqnarray}
 \displaystyle
  \Delta C^{\mu} &=&-p_3^\mu\, \Delta C_0 +\frac{1}{2}\,\frac{\partial}{\partial p_{2\mu}}\Delta C_1,\nonumber\\
  \Delta C^{\mu\nu} &=&p_3^\mu\, p_3^\nu\, \Delta C_0 -\frac{1}{2}\left(g^{\mu\nu}+p_3^\mu\frac{\partial}{\partial p_{2\nu}}+p_3^\nu\frac{\partial}{\partial p_{2\mu}}\right)\Delta C_1 +\frac{1}{4}\frac{\partial}{\partial p_{2\mu}}\frac{\partial}{\partial p_{2\nu}} \Delta C_2\,,\nonumber\\ 
  \Delta C_1  &=& \Delta C_0 \left(f_0 \rightarrow f_1\right)\,,\nonumber\\
  \Delta C_2 &=&\Delta C_0\left(f_0 \rightarrow f_2\right).
   \label{C1C2}
   \end{eqnarray}
Then, computing $\Delta C_1$ and $\Delta C_2$ integrals in a straightforward manner, we find
\begin{eqnarray}
\!\!\!\!\!\!\!\!\!\!\!\! \Delta C_1 &=& -\frac{\alpha +\beta }{2 p_1^2}\, \Delta C_0 +\frac{2\pi i}{p_1^2}\,\sqrt{\lambda(p_1^2,m_1^2,m_2^2)}\;\log\left(\frac{\alpha-\beta}{2\,p_1^2\, e}\right)\Theta\left(p_1^2-(m_1+m_2)^2\right)\,,\nonumber\\
\!\!\!\!\!\!\!\!\!\!\!\! \Delta C_2 &=& \frac{(\alpha +\beta)^2}{8p_1^4}\, \Delta C_0-\frac{\pi \alpha i}{p_1^4}\,\sqrt{\lambda(p_1^2,m_1^2,m_2^2)}\;\log\left (\frac{\alpha-\beta}{2\,p_1^2\,e^{3/2}}\right)\Theta\left(p_1^2-(m_1+m_2)^2\right)
\end{eqnarray}
Here $e$ is the Napier's constant. Next, we take the derivatives of $\Delta C_1$ and $\Delta C_2$ with respect to $p_{2,3}$ and plug them into Eq.~(\ref{C1C2}). After some lengthy algebra, we get 
\begin{eqnarray}
\displaystyle
 \!\!\!\Delta C^\mu\!\! &=&\!\! \frac{2\pi i\, \Theta(p_1^2-(m_1+m_2)^2)}{\sqrt{\lambda(p_1^2,p_2^2,p_3^2)}}\,\left[\log\left(\frac{\alpha+\beta}{\alpha-\beta}\right)p_3^\mu
 -\left((u-\alpha v)\,\log\left(\frac{\alpha+\beta}{\alpha-\beta}\right)+2\beta v\right)\frac{p_2^\mu}{2\,p_1^2}\right],\nonumber\\
\!\!\Delta C^{\mu\nu}\!\! &=&\!\! \frac{2\pi i\, \Theta(p_1^2-(m_1+m_2)^2)}{p_1^2 \sqrt{\lambda(p_1^2,p_2^2,p_3^2)}}\,\left(\frac{g^{\mu\nu}}{2} \Delta C_g + \frac{p_2^\mu\, p_2^\nu}{8\,p_1^2}\, \Delta C_{22} + (p_2^\mu\, p_3^\nu+p_2^\nu\, p_3^\mu)\,\Delta C_{23}+p_3^\mu\, p_3^\nu\, \Delta C_{33} \right),\nonumber\\
 \!\!\!\Delta C_g\!\! &=&\!\! \frac{\alpha\,\beta\, v}{4\,p_1^2}-
  (\alpha+\beta)\left(\frac{1}{2}-\frac{u}{4\,p_1^2}+\frac{(\alpha-\beta)v}{8\,p_1^2}\right)\log\left(\frac{\alpha+\beta}{\alpha-\beta}\right)
 - \beta\left(1-\frac{u}{2\,p_1^2}\right)\log\left(\frac{\alpha-\beta}{2\,p_1^2\,e}\right),\nonumber\\
 \!\!\!\Delta C_{22}\!\! &=&\!\! \left[(\alpha^2-\beta^2)\left(\frac{1}{\lambda(p_1^2,p_2^2,p_3^2)}-v^2\right)-2(u-\alpha\,v)^2\right] \log\left(\frac{\alpha+\beta}{\alpha-\beta}\right)+8\beta v(u-\alpha\,v)\nonumber\\
 &&+2\alpha\beta\left(\frac{1}{\lambda(p_1^2,p_2^2,p_3^2)}+v^2\right),\nonumber\\
\!\!\!\Delta C_{23}\!\! &=&\!\beta v + \frac{u-\alpha\,v}{2}\,\log\left(\frac{\alpha+\beta}{\alpha-\beta}\right),\nonumber\\
 \!\!\!\Delta C_{33}\!\! &=&-p_1^2\,\log\left(\frac{\alpha+\beta}{\alpha-\beta}\right)\,,
\end{eqnarray} 
where $u=p_1^2+m_1^2-m_2^2$ and $v=(p_1^2-p_2^2+p_3^2)/\lambda(p_1^2,p_2^2,p_3^2)$ and $g^{\mu\nu}=(+,-,-,-)$ is the metric tensor. 

It remains only to determine the coefficients, like the ones defined in Eq.~(\ref{Yn}), by comparing the actual matrix elements for the diagrams in Fig.~\ref{fig:relevant}. The diagrams in  Fig.~\ref{fig:relevant} and the corresponding matrix elements are generated with the software {\tt FeynArts} \cite{Hahn:2000jm} and then we do the rest of the calculation with our own code.


\begin{thebibliography}{99}
%
\bibitem{Yao:2006px}
  W.~M.~Yao {\it et al.}  [Particle Data Group],
  J.\ Phys.\ G {\bf 33}, 1 (2006).

  		  
  		  
\bibitem{Aubert:2001fg} B.~Aubert {\it et al.}  [BABAR Collaboration],
  			Phys.\ Rev.\  D {\bf 65}, 091101 (2002);
  			K.~Abe {\it et al.}  [Belle Collaboration],
                        Phys.\ Rev.\ Lett.\  {\bf 87}, 091802 (2001).
			

\bibitem{Buchmuller:1982ye} W.~Buchmuller and D.~Wyler,
                            Phys.\ Lett.\ B {\bf 121}, 321 (1983);
                            J.~Polchinski and M.~B.~Wise,
                            Phys.\ Lett.\ B {\bf 125}, 393 (1983);
                            A.~De Rujula, M.~B.~Gavela, O.~Pene and F.~J.~Vegas,
                            Phys.\ Lett.\ B {\bf 245}, 640 (1990).

                            
\bibitem{Hao:2007qv}
  		S.~Hao, M.~Wen-Gan, Z.~Ren-You, G.~Lei, H.~Liang and J.~Yi,
  		Phys.\ Rev.\  D {\bf 75}, 095006 (2007).


\bibitem{Coniavitis:2007me}
  		E.~Coniavitis and A.~Ferrari,
  		Phys.\ Rev.\  D {\bf 75}, 015004 (2007).




\bibitem{Weiglein:2004hn}
  		G.~Weiglein {\it et al.}  [LHC/LC Study Group],
  		Phys.\ Rept.\  {\bf 426}, 47 (2006).


\bibitem{Demir:2003bv} D.~A.~Demir,
                       Phys.\ Lett.\  B {\bf 571}, 193 (2003).


\bibitem{Christova:2002ke} E.~Christova, H.~Eberl, W.~Majerotto and S.~Kraml,
                           Nucl.\ Phys.\  B {\bf 639}, 263 (2002)
                           [Erratum-ibid.\  B {\bf 647}, 359 (2002)].
                           			  
              
\bibitem{Christova:2002sw} E.~Christova, H.~Eberl, W.~Majerotto and S.~Kraml,
                           JHEP {\bf 0212}, 021 (2002);
                           E.~Christova, E.~Ginina and M.~Stoilov,
                           JHEP {\bf 0311}, 027 (2003).
                           

\bibitem{Christova:2006fb} E.~Christova, H.~Eberl, E.~Ginina and W.~Majerotto,
                           JHEP {\bf 0702}, 075 (2007).

\bibitem{Eilam:1991yv} G.~Eilam, J.~L.~Hewett and A.~Soni,
                       Phys.\ Rev.\ Lett.\  {\bf 67}, 1979 (1991).
                       
 \bibitem{Besmer:2001cj}
 		 T.~Besmer, C.~Greub and T.~Hurth,
  		Nucl.\ Phys.\ B {\bf 609}, 359 (2001);
  		D.~A.~Demir,
  		Phys.\ Lett.\ B {\bf 571}, 193 (2003);
  		A.~M.~Curiel, M.~J.~Herrero and D.~Temes,
  		Phys.\ Rev.\ D {\bf 67}, 075008 (2003);
  		J.~J.~Liu, C.~S.~Li, L.~L.~Yang and L.~G.~Jin,
  		Nucl.\ Phys.\ B {\bf 705}, 3 (2005).

\bibitem{Chankowski:2005jh}
		  P.~H.~Chankowski, O.~Lebedev and S.~Pokorski,
		  Nucl.\ Phys.\  B {\bf 717}, 190 (2005).

\bibitem{Frank:2006ku} M.~Frank and I.~Turan,
                       Phys.\ Rev.\  D {\bf 74}, 073014 (2006).

\bibitem{Fischler:1992ha} W.~Fischler, S.~Paban and S.~D.~Thomas,
                          Phys.\ Lett.\  B {\bf 289}, 373 (199);
                          S.~M.~Barr,
  			  Int.\ J.\ Mod.\ Phys.\  A {\bf 8}, 209 (1993);
  			  V.~D.~Barger, T.~Falk, T.~Han, J.~Jiang, T.~Li and T.~Plehn,
                           Phys.\ Rev.\  D {\bf 64}, 056007 (2001);  
  			  C.~A.~Baker {\it et al.},
                          Phys.\ Rev.\ Lett.\  {\bf 97}, 131801 (2006).  
                          		      
\bibitem{Demir:2003js} D.~A.~Demir, O.~Lebedev, K.~A.~Olive, M.~Pospelov and A.~Ritz,
                       Nucl.\ Phys.\  B {\bf 680}, 339 (2004);
                       D.~A.~Demir and M.~B.~Voloshin,
                       Phys.\ Rev.\  D {\bf 63}, 115011 (2001).
                       
  
\bibitem{Gabbiani:1996hi} F.~Gabbiani, E.~Gabrielli, A.~Masiero and L.~Silvestrini,
                          Nucl.\ Phys.\ {\bf B477}, 321 (1996);
                          M.~Misiak, S.~Pokorski and J.~Rosiek,
                          Adv.\ Ser.\ Direct.\ High Energy Phys.\  {\bf 15}, 795 (1998);
                          E.~Lunghi, A.~Masiero, I.~Scimemi and L.~Silvestrini,
			  Nucl.\ Phys.\  B {\bf 568}, 120 (2000);
                          M.~Ciuchini, E.~Franco, A.~Masiero, L.~Silvestrini,
                          Phys.\ Rev.\ {\bf D67}, 075016 (2003) [Erratum ibid.\ {\bf D68},
                          079901 (2003).
                          
			  

\bibitem{Atwood:2000tu} D.~Atwood, S.~Bar-Shalom, G.~Eilam and A.~Soni,
  			Phys.\ Rept.\  {\bf 347}, 1 (2001).
			                     

                       	                     
\bibitem{Bi:1999is} X.~J.~Bi and Y.~B.~Dai,
                    Eur.\ Phys.\ J.\  C {\bf 12}, 125 (2000).

\bibitem{Passarino:1978jh} G.~Passarino and M.~J.~G.~Veltman,
                           Nucl.\ Phys.\  B {\bf 160}, 151 (1979).
                                	                     
\bibitem{Cutkosky:1960sp} R.~E.~Cutkosky,
                          J.\ Math.\ Phys.\  {\bf 1}, 429 (1960);
                          G.~'t Hooft and M.~Veltman,
  			  {\it Diagrammar},~CERN-73-09;
  			  M.~E.~Peskin and D.~V.~Schroeder,
                          {\it An Introduction To Quantum Field Theory},
			  Addison-Wesley, USA (1995).	
                          
                                   	                     
\bibitem{Bauberger:1994hx} S.~Bauberger and M.~Bohm,
                           Nucl.\ Phys.\  B {\bf 445}, 25 (1995);
                           S.~Bauberger, M.~Bohm, G.~Weiglein, F.~A.~Berends and M.~Buza,
                           Nucl.\ Phys.\ Proc.\ Suppl.\  {\bf 37B}, 95 (1994).
                            

\bibitem{Scharf:1996zi} G.~Scharf,
  			{\it Finite quantum electrodynamics: The Causal approach},
                        Springer, Berlin (1995). 
                                  
\bibitem{Hahn:2000jm} T.~Hahn and M.~Perez-Victoria,
                      Comput.\ Phys.\ Commun.\  {\bf 118}, 153 (1999);
                      T.~Hahn,
                      Nucl.\ Phys.\ Proc.\ Suppl.\ {\bf 89}, 231 (2000);
                      T.~Hahn,
                      Comput.\ Phys.\ Commun.\ {\bf 140}, 418 (2001);
                      T.~ Hahn and C.~Schappacher,
                      Comput.\ Phys.\ Commun.\ {bf 143}, 54 (2002).

\bibitem{FeynHiggs} S.~Heinemeyer, W.~Hollik and G.~Weiglein,
                    Comput.\ Phys.\ Comm.\ {\bf 124} 76 (2000);
                    S.~Heinemeyer, W.~Hollik and G.~Weiglein,
                     Eur.\ Phys.\ J.\ {\bf C 9} 343 (1999);
                    G.~Degrassi, S.~Heinemeyer, W.~Hollik,
                    P.~Slavich and G.~Weiglein,
                    Eur.\ Phys.\ J.\ {\bf C 28} 133 (2003);
                    T.~Hahn, W.~Hollik, S.~Heinemeyer and G.~Weiglein,
                    {\it In the Proceedings of 2005 International Linear Collider Workshop (LCWS 2005), Stanford, California, 18-22 Mar 2005, pp 0106};
		    M.~Frank, T.~Hahn, S.~Heinemeyer, W.~Hollik, H.~Rzehak and G.~Weiglein,
                    JHEP {\bf 0702}, 047 (2007).
                      
\bibitem{Djouadi:1995gv} A.~Djouadi, J.~Kalinowski and P.~M.~Zerwas,
                         Z.\ Phys.\  C {\bf 70}, 435 (1996).                         
			                     
\end{thebibliography}

\end{document}